\documentclass[12pt]{article}
\usepackage{amsthm,amsmath,amsfonts,amssymb}
\usepackage[colorlinks,citecolor=blue,urlcolor=blue]{hyperref}
\usepackage{graphicx}
\graphicspath{{figure/}} 
\usepackage{float}
\usepackage{tikz}
\usepackage{bm}
\usepackage{booktabs}
\usepackage{multirow}
\usepackage{dsfont}
\usepackage{subcaption}
\usepackage{enumerate}
\usepackage{natbib}
\usepackage{url} 

\theoremstyle{plain}
\newtheorem{theorem}{Theorem}
\newtheorem{lemma}{Lemma}
\newtheorem{prop}{Proposition}
\newtheorem{corollary}{Corollary}
\theoremstyle{remark}

\newtheorem{assum}{Assumption}
\newcommand{\E}{\mathbb{E}}
\newcommand{\K}{\mathrm{K}}
\newcommand{\prob}{\mathbb{P}}
\newcommand{\reall}{\mathbb{R}}
\newcommand{\sign}{\mathrm{sign}}

\newcommand{\wsp}{\mathcal{W}}

\newcommand{\tdomain}{\mathcal{D}}
\newcommand{\idomain}{I}

\newcommand{\mcalT}{\mathcal{T}}
\newcommand{\wdomain}{\mathcal{S}}
\newcommand{\dwsp}{\wsp} 
\def\diff{{\rm d}}
\def\argmin#1{\mathrel{\mathop{\arg\min}\limits_{#1}}}

\newcommand{\bea}{\begin{eqnarray*}}
\newcommand{\eea}{\end{eqnarray*}}
\newcommand{\be}{\begin{eqnarray}}
\newcommand{\ee}{\end{eqnarray}}
\newcommand{\bsp}{\begin{split}}
\newcommand{\esp}{\end{split}}
\newcommand{\ed}{\end{document}}
\newcommand{\no}{\noindent}

\newcommand{\btab}{\begin{tabular}}
\newcommand{\etab}{\end{tabular}}
\newcommand{\bc}{\begin{center}}
\newcommand{\ec}{\end{center}}

\newcommand{\bi}{\begin{itemize}}
\newcommand{\ei}{\end{itemize}}
\newcommand{\bfi}{\begin{figure}}
\newcommand{\efi}{\end{figure}}
\newcommand{\ben}{\begin{enumerate}}
\newcommand{\een}{\end{enumerate}}
\newcommand{\bdes}{\begin{description}}
\newcommand{\edes}{\end{description}}
\newcommand{\bay}{\begin{array}}
\newcommand{\eay}{\end{array}}

\DeclareMathOperator*{\arginf}{\arg\inf}
\def\red{\textcolor{red}}

\def\bco{\iffalse}

\def\cp{\citep}

\def\F{Fr\'echet}

\newcommand{\bass}{\begin{assumption}}
\newcommand{\eass}{\end{assumption}}
\newcommand{\bthm}{\begin{theorem}}
\newcommand{\ethm}{\end{theorem}}
\newcommand{\blem}{\begin{lemma}}
\newcommand{\elem}{\end{lemma}}

\newcommand{\single}{\renewcommand{\baselinestretch}{1.2}\normalsize}
\newcommand{\double}{\renewcommand{\baselinestretch}{1.6}\normalsize}

\def\bco{\iffalse}

\def\cp{\citep}

\def\F{Fr\'echet }

\newif\ifsupp
\suppfalse 
\supptrue

\newcommand{\suppcontent}[1]{%
    \ifsupp
    
    \else
    
       \red{===============================================}
       
        #1 
        
       \red{===============================================}
       
    \fi
}

\newcommand{\blind}{1}

\begin{document}

\def\spacingset#1{\renewcommand{\baselinestretch}%
{#1}\small\normalsize} \spacingset{1}


\if1\blind
{
  \title{\bf Functional Principal Component Analysis for  Distribution-Valued Processes}
  \author{Hang Zhou and Hans-Georg M\"{u}ller
  \thanks{This research was supported in part by NSF grants DMS-2014626 and DMS-2310450.} \\
\\
    Department of Statistics, University of California, Davis, CA 95616\\
    }
  \maketitle
} \fi

\if0\blind
{
  \bigskip
  \bigskip
  \bigskip
  \begin{center}
    {\LARGE\bf Functional Principal Component Analysis for 
    Distribution-Valued Processes}
\end{center}
  \medskip
} \fi

\bigskip
\begin{abstract}
We develop statistical models for samples of distribution-valued stochastic processes featuring time-indexed univariate distributions, with  emphasis on functional principal component analysis. The  proposed model presents an intrinsic rather than transformation-based approach. The starting point is a  transport process representation for distribution-valued processes under the Wasserstein metric. Substituting transports for distributions addresses the challenge of centering distribution-valued processes and  leads to a useful and interpretable decomposition  of each realized process into a  process-specific single transport and a real-valued trajectory. This representation makes it possible 
to utilize  a scalar multiplication operation for transports and facilitates not only functional principal component analysis  but also to introduce a latent  Gaussian process. This Gaussian process proves  especially useful for the case where the distribution-valued processes are only observed on a sparse grid of time points, establishing an approach for  longitudinal distribution-valued data. We study  the convergence of the key components of this novel representation to their population targets and demonstrate the practical utility of the proposed approach through simulations and several data illustrations.
\end{abstract}

\noindent%
\vfill

\newpage
\spacingset{1.2} 

\section{Introduction}

Functional data are samples of realizations of square integrable scalar or vector-valued functions that have been extensively studied 
 \citep{ramsay2005,hsing2015, mull:16:3,Kokoszka2017}. The  restriction to the realm of Euclidean space-valued functions that also encompasses  Hilbert-space
valued functional data, i.e., function-valued stochastic processes  \cp{mull:12:3,mull:17:4}, 
  is an essential feature of functional data, but 
 proves  too restrictive as new complex non-Euclidean data types are emerging. A previous very general model for the case of a  metric-space valued process for which one observes a sample of realizations  \citep{dubey2020functional} includes distribution-valued processes as a special case. 
 The general  framework developed in this previous approach 
 utilizes a notion of metric covariance and includes a certain kind of functional principal component analysis  for general metric space-valued processes by using Fr\'echet integrals \citep{petersen2016frechet} and is limited to the case of  fully observed metric space-valued functional data, where it  is assumed that $X_{i}(t)$ is known for all $t$ in the time domain and  cannot be extended to the case of sparsely sampled processes. 
 The generality of this framework also means that the provided tools are somewhat limited, 
  due to  the lack of algebraic structure or geodesics in general metric spaces.  

A narrower class of non-Euclidean valued processes, where one has more structure than in the general metric case,   are random object-valued processes that take values on Riemannian manifolds. This special class of processes,  exemplified by repeatedly observed flight paths on Earth, has been studied through the application of  Riemannian log maps, where  the Riemannian random objects at fixed arguments are mapped to the linear tangent spac e at  a reference point. One can then  perform subsequent analysis  on the linear spaces of the log processes  \citep{lin2019, dai2021modeling}, which are situated in linear tangent spaces, where one can take advantage of the usual Euclidean geometry. 

Our goal in this paper is to develop models and analysis tools for a specific yet important class of random object-valued stochastic processes: those where the  time-indexed objects are univariate distributions.
The argument of the process is referred to as time in the following but could be any scalar that varies over an interval. Distribution-valued stochastic processes are encountered in various complex applications that include  country-specific age-at-death distributions, fertility distributions or income distributions over calendar years for a sample of countries. The basic starting point throughout is that one has 
an i.i.d. sample of realizations of such processes. The statistical modeling of distribution-valued  processes is an essential yet still missing tool for the  emerging field of distributional data analysis \citep{pete:22}, while various modeling approaches for distributional  regression and distributional time series  have been studied  recently
  \cp{koko:19, ghod:21,chen2021wasserstein, mull:21:3}. 

We aim for intrinsic modeling of distributions rather than  extrinsic approaches,  where one first transforms distributions to a linear space \citep{scealy2011regression,petersen2016functional,zhang2022nonlinear,chen2021wasserstein} and then applies functional data analysis methodology in this linear space and finally transforms back to the metric space.  These transformation approaches  are somewhat arbitrary  and have various downsides.  For example,
\cite{petersen2016functional} proposed a family  of global transformations of distributions to a Hilbert space,  with the most prominent representative being the  log quantile density transformation, however these transformations are  metric-distorting. On the other hand,  log transformations to tangent bundles are isometric but the inverse exp maps are  not well defined 
on the entire tangent space,   which causes major problems especially for principal component analysis; various  ad-hoc solutions  have been proposed to address these issues \citep{bigo:17, pego:22, chen2021wasserstein}. 

An issue that is of additional practical relevance and theoretical interest is that available
observations typically are not available continuously  in time but only  at discrete time points. A common model assumption for such sparse designs is   that one observes the process only at a few randomly located timepoints. 
These considerations  motivate our goal to develop a comprehensive intrinsic model for distribution-valued processes where the processes may be fully or only partially observed. 
Throughout we work with the 2-Wasserstein metric $d_{W,2}$ and optimal transports, which move distributions along geodesics.  The challenge of intrinsic modeling is that the Wasserstein space of distributions does not have a linear or vector space structure. This challenge can be addressed by making use of rudimentary algebraic operations on the space of optimal transports \citep{mull:21:3}.   From the outset we aim to deal with centered processes. Since no subtraction exists in the Wasserstein space, the centering of distribution-valued processes is achieved by substituting transport processes for distributional processes:  For each time argument the distributions that constitute the values of a distributional process at a fixed time $t$ are replaced by  transports from the barycenter (Fr\'echet mean) of the process at $t$  to the distribution that corresponds to the value of the process at time $t$.  These transports are well defined if one adopts the  Wasserstein metric. Their Fr\'echet mean is the identity transport, i.e., these transports are centered. In the following we will therefore refer to the processes that we study as transport processes rather than distributional processes.

We motivate the  proposed methodology with the modeling of age-at-death distributonal processes as observed for a sample of countries. Other pertinent examples include the distributions of price fluctuations in finance/economics/housing  \citep{chen2021wasserstein, mull:21:3} and the distributions of signal strength in functional magnetic resonance imaging studies \citep{zhou2021intrinsic}. 
For our study of  stochastic transport processes we introduce  representations
$$
	T(t) = g(Z(t))\odot T_{0},
$$
where $Z(t)$ is a $\reall$-valued random process, $g$ is a bijective function that maps $\reall$ to {$(-1,1)$} and  $T_{0}$ is a single random transport that is a summary characteristic for each realization of the transport process.  Here  $\odot$ is a multiplication operation by which  a transport is multiplied with a scalar \citep{mull:21:3}. By construction, $g(Z(t))\odot T_{0}$ lies on the extended geodesic that passes through $T_{0}$. 
We develop a predictor for each individual $T_i(t)$ based on observations obtained at discrete time points and  establish asymptotic convergence rates for the components of the model for both densely and sparsely sampled distributional processes.  These are  novel even for  classical real-valued functional data. 

The rest of this paper is organized as follows. Section \ref{sec:WTspace} provides a brief introduction to the  transport space.  The proposed methodology and transport model are introduced in Section \ref{sec:meth} and the theoretical results are presented in Section \ref{sec:theo}.   Section \ref{sec:simu} contains numerical studies for synthetic data. We illustrate the method in Section \ref{sec:appli} with human mortality and glucose density data. Proofs and auxiliary results  are provided in the Supplement.

\section{Transforming distribution-valued data to  transports}\label{sec:WTspace}
Let $\dwsp$ be the set of finite second moment probability measures on the closed interval  $\wdomain\subset \reall $, 
\be \label{W}	\dwsp=\left\{\mu\in\mathcal{P}(\wdomain):\,\, \int_{\wdomain}|x|^2\diff \mu(x)<\infty \right\},
\ee
where $\mathcal{P}(\wdomain)$ is the set of all probability measures on $\wdomain$. The $p$-Wasserstein distance $d_{W,p}(\cdot, \cdot)$ between two measures $\mu, \nu \in \dwsp$ is 
\begin{equation}\label{eq:def-kan}
	d_{W,p}({\mu},{\nu}):=\inf\bigg\{\left(\int_{\wdomain^2}|x_{1}-x_{2}|^p\diff \Gamma  (x_1,x_2)\right)^{1/p}: \,\, \Gamma  \in \Gamma  (\mu,\nu)\bigg\}\quad \text{ for }p>0,
\end{equation}
where $\Gamma(\mu, \nu)$ is the set of joint probability measures on $\wdomain^2$ with $\mu$ and $\nu$ as marginal measures.  The Wasserstein space $(\dwsp, d_{W,p})$ is a separable and complete metric space 
\citep{ambrosio2008,villani2009optimal}. Here we assume $\mathcal{S}=[0,1]$ without loss of generality to simply the notation.
Given two probability measures $\mu,\nu \in \dwsp$, the optimal transport from $\mu$ to $\nu$ is the 
map $T:\wdomain \to \wdomain$ that minimizes the transport cost, 
\begin{equation}\label{eq:def-monge}
		\arginf_{T\in\mathcal{T}} \bigg\{\left( \int_{\wdomain}|T(u)-u|^p\diff \mu(u)\right)^{1/p},\text{ such that } T\#\mu=\nu\bigg\},
\end{equation}
where $\mathcal{T}=\{T:\wdomain\mapsto\wdomain| T(0)=0,\, T(1)=1,T \, \text{ is non-decreasing} \}$ is the transport space and $T\#\mu$ is the push-forward measure of $\mu$, 
defined as $(T\#\mu)(A) = \mu\{x \in \wdomain \mid T(x) \in A\}$ for all $A$ in the Borel algebra of $\wdomain$. This optimization problem, also known as the Monge problem, is a relaxation of the Kantorovich problem \eqref{eq:def-kan}. If $\mu$ is absolutely continuous with respect to the Lebesgue measure, then problems \eqref{eq:def-kan} and \eqref{eq:def-monge} are equivalent and have a unique solution $T(u) = F_{\nu}^{-1} \circ F_{\mu}(u)$ for $p=2$, where $F_{\mu}$ and $F_{\nu}^{-1}$ are the cumulative distribution  and quantile functions of $\mu$ and $\nu$, respectively \citep{gangbo1996geometry}. 

We will demonstrate that optimal transport is instrumental to overcome the challenge of the non-linearity of the Wasserstein space, specifically  the absence of the subtraction operation, 
and thus  to extend functional principal component analysis to $\dwsp$-valued functional data. Indeed, optimal transport between two measures can be interpreted as the equivalent of the subtraction operation in linear spaces, where the starting measure is ``subtracted'' from the measure resulting from the transport.  For  a distribution-valued process  $X(t)$ with random distributions on domain $\mathcal{S}$  where $t \in \tdomain$ for  a closed interval in $\mathbb{R}$, the cross-sectional Fr\'echet mean of $X(t)$ at each $t$ is $$\mu_{\oplus,2}(t)=\text{argmin}_{\omega\in\dwsp}\E d_{W,2}^{2}(X(t),\omega).$$
We then define the (optimal) transport process $T(\cdot)$, where  
$T(t)$ represents the optimal transport from $\mu_{\oplus,2}(t)$ to $X(t)$, $\mu_{\oplus,2}(t)$ serves as the mean, and the transport $T(t)$ from $\mu_{\oplus,2}(t)$ 
to $X(t)$
quantifies the difference between $X(t)$ and 
$\mu_{\oplus,2}(t)$ for each $t\in \tdomain$ under the Wasserstein metric. Here  $T(t)$ 
is akin to a centered process, where the Fr\'echet  mean of $T(t)$ is the identity transport and thus the null element for all $t$.  

An illustrative example is in Section \ref{sec:appli}, where the realized processes $X_i(t)$ are the age-at-death distributions of $n=33$ countries with time being the calendar year. 
Then  $T_i(t)$ reflects how the age-at-death distribution of a specific country differs from the \F mean  of all 33 countries at calendar year $t$.

It is thus advantageous to use the transport space $\mathcal{T}$ for the statistical modeling of Wasserstein space-valued stochastic processes.  Note that $\mathcal{T}$ is a closed subset of $\mathcal{L}^{p}(\wdomain)=\{f:\wdomain\mapsto \reall|\,\, \|f\|_{p}<\infty  \}$, where $\|f\|_{p}=(\int_{\wdomain}|f(x)|^{p}\diff x )^{1/p}$ is the usual $\mathcal{L}^{p}$-norm. Hence, $(\mathcal{T},d_{W,p})$ is a complete metric space with $d_{W,p}(T_{1},T_{2})=(\int_{\wdomain}|T_{1}(x)-T_{2}(x)|^{p}\diff x )^{1/p}$, endowed with the norm $\|T\|_{p}=(\int_{\wdomain}|T(x)|^{p}\diff x )^{1/p}$. Proposition S.2 in the Supplement demonstrates that the Wasserstein space and the transport space are isometric.
\suppcontent{
\begin{prop}\label{prop:iso}
	There exists an isometric map $\mathfrak{M}:\dwsp\mapsto \mcalT$ between $(\dwsp,d_{W,2} )$ and $(\mcalT,d_{W,2} )$ given by
\begin{equation}\label{def:eq-isoM}
\mathfrak{M}(\mu)=F_{\mu}^{-1}\circ F_{\wdomain} \text{ and } \mathfrak{M}^{-1}(T)=T\# \mathrm{Unif}{\wdomain},
\end{equation}
for all $\mu\in\dwsp$ and $T\in\mcalT$, where $\mathrm{Unif}_{\wdomain}$ is the uniform distribution on $\wdomain$ and $F_{\wdomain}$ is the cumulative distribution function of $\mathrm{Unif}_{\wdomain}$.
\end{prop}

\begin{figure}

\tikzset{every picture/.style={line width=0.75pt}} 

\begin{tikzpicture}[x=0.75pt,y=0.75pt,yscale=-1,xscale=1]

\draw [color={rgb, 255:red, 126; green, 211; blue, 33 }  ,draw opacity=1 ]   (194.23,191.58) -- (155,138) ;
\draw [shift={(196,194)}, rotate = 233.79] [fill={rgb, 255:red, 126; green, 211; blue, 33 }  ,fill opacity=1 ][line width=0.08]  [draw opacity=0] (8.93,-4.29) -- (0,0) -- (8.93,4.29) -- cycle    ;
\draw   (70,67) .. controls (59,2) and (203,35) .. (233,67) .. controls (263,99) and (267,198) .. (249,219) .. controls (231,240) and (111,251) .. (76,211) .. controls (41,171) and (81,132) .. (70,67) -- cycle ;
\draw   (399,75) .. controls (420,11) and (565,18) .. (564,73) .. controls (563,128) and (612,199) .. (579,224) .. controls (546,249) and (436,256) .. (407,217) .. controls (378,178) and (378,139) .. (399,75) -- cycle ;
\draw [color={rgb, 255:red, 74; green, 144; blue, 226 }  ,draw opacity=1 ]   (155,138) -- (155,214) ;
\draw [shift={(155,217)}, rotate = 270] [fill={rgb, 255:red, 74; green, 144; blue, 226 }  ,fill opacity=1 ][line width=0.08]  [draw opacity=0] (8.93,-4.29) -- (0,0) -- (8.93,4.29) -- cycle    ;
\draw [shift={(155,138)}, rotate = 90] [color={rgb, 255:red, 74; green, 144; blue, 226 }  ,draw opacity=1 ][fill={rgb, 255:red, 74; green, 144; blue, 226 }  ,fill opacity=1 ][line width=0.75]      (0, 0) circle [x radius= 3.35, y radius= 3.35]   ;
\draw [color={rgb, 255:red, 208; green, 2; blue, 27 }  ,draw opacity=1 ]   (155,138) -- (115.77,84.42) ;
\draw [shift={(114,82)}, rotate = 53.79] [fill={rgb, 255:red, 208; green, 2; blue, 27 }  ,fill opacity=1 ][line width=0.08]  [draw opacity=0] (8.93,-4.29) -- (0,0) -- (8.93,4.29) -- cycle    ;
\draw [color={rgb, 255:red, 74; green, 144; blue, 226 }  ,draw opacity=1 ]   (114,82) -- (114,158) ;
\draw [shift={(114,161)}, rotate = 270] [fill={rgb, 255:red, 74; green, 144; blue, 226 }  ,fill opacity=1 ][line width=0.08]  [draw opacity=0] (8.93,-4.29) -- (0,0) -- (8.93,4.29) -- cycle    ;
\draw [shift={(114,82)}, rotate = 90] [color={rgb, 255:red, 74; green, 144; blue, 226 }  ,draw opacity=1 ][fill={rgb, 255:red, 74; green, 144; blue, 226 }  ,fill opacity=1 ][line width=0.75]      (0, 0) circle [x radius= 3.35, y radius= 3.35]   ;
\draw    (114,161) ;
\draw [shift={(114,161)}, rotate = 0] [color={rgb, 255:red, 0; green, 0; blue, 0 }  ][fill={rgb, 255:red, 0; green, 0; blue, 0 }  ][line width=0.75]      (0, 0) circle [x radius= 3.35, y radius= 3.35]   ;
\draw    (155,217) ;
\draw [shift={(155,217)}, rotate = 0] [color={rgb, 255:red, 0; green, 0; blue, 0 }  ][fill={rgb, 255:red, 0; green, 0; blue, 0 }  ][line width=0.75]      (0, 0) circle [x radius= 3.35, y radius= 3.35]   ;
\draw  [dash pattern={on 4.5pt off 4.5pt}]  (114,82) -- (473,81.01) ;
\draw [shift={(475,81)}, rotate = 179.84] [color={rgb, 255:red, 0; green, 0; blue, 0 }  ][line width=0.75]    (10.93,-3.29) .. controls (6.95,-1.4) and (3.31,-0.3) .. (0,0) .. controls (3.31,0.3) and (6.95,1.4) .. (10.93,3.29)   ;
\draw    (487,142) ;
\draw    (487,142) ;
\draw [shift={(487,142)}, rotate = 0] [color={rgb, 255:red, 0; green, 0; blue, 0 }  ][fill={rgb, 255:red, 0; green, 0; blue, 0 }  ][line width=0.75]      (0, 0) circle [x radius= 3.35, y radius= 3.35]   ;
\draw  [dash pattern={on 4.5pt off 4.5pt}]  (157,216.99) -- (448,215) ;
\draw [shift={(155,217)}, rotate = 359.61] [color={rgb, 255:red, 0; green, 0; blue, 0 }  ][line width=0.75]    (10.93,-3.29) .. controls (6.95,-1.4) and (3.31,-0.3) .. (0,0) .. controls (3.31,0.3) and (6.95,1.4) .. (10.93,3.29)   ;
\draw [color={rgb, 255:red, 74; green, 144; blue, 226 }  ,draw opacity=1 ]   (448,215) ;
\draw [shift={(448,215)}, rotate = 0] [color={rgb, 255:red, 74; green, 144; blue, 226 }  ,draw opacity=1 ][fill={rgb, 255:red, 74; green, 144; blue, 226 }  ,fill opacity=1 ][line width=0.75]      (0, 0) circle [x radius= 3.35, y radius= 3.35]   ;
\draw    (155,138) ;
\draw [shift={(155,138)}, rotate = 0] [color={rgb, 255:red, 0; green, 0; blue, 0 }  ][fill={rgb, 255:red, 0; green, 0; blue, 0 }  ][line width=0.75]      (0, 0) circle [x radius= 3.35, y radius= 3.35]   ;
\draw    (114,82) ;
\draw [shift={(114,82)}, rotate = 0] [color={rgb, 255:red, 0; green, 0; blue, 0 }  ][fill={rgb, 255:red, 0; green, 0; blue, 0 }  ][line width=0.75]      (0, 0) circle [x radius= 3.35, y radius= 3.35]   ;
\draw    (475,81) .. controls (490,115) and (469,99) .. (487,142) ;
\draw    (487,142) .. controls (502,176) and (481,160) .. (499,203) ;
\draw [color={rgb, 255:red, 208; green, 2; blue, 27 }  ,draw opacity=1 ]   (475,81) ;
\draw [shift={(475,81)}, rotate = 0] [color={rgb, 255:red, 208; green, 2; blue, 27 }  ,draw opacity=1 ][fill={rgb, 255:red, 208; green, 2; blue, 27 }  ,fill opacity=1 ][line width=0.75]      (0, 0) circle [x radius= 3.35, y radius= 3.35]   ;
\draw [color={rgb, 255:red, 126; green, 211; blue, 33 }  ,draw opacity=1 ]   (499,203) ;
\draw [shift={(499,203)}, rotate = 0] [color={rgb, 255:red, 126; green, 211; blue, 33 }  ,draw opacity=1 ][fill={rgb, 255:red, 126; green, 211; blue, 33 }  ,fill opacity=1 ][line width=0.75]      (0, 0) circle [x radius= 3.35, y radius= 3.35]   ;
\draw    (196,194) ;
\draw [shift={(196,194)}, rotate = 0] [color={rgb, 255:red, 0; green, 0; blue, 0 }  ][fill={rgb, 255:red, 0; green, 0; blue, 0 }  ][line width=0.75]      (0, 0) circle [x radius= 3.35, y radius= 3.35]   ;

\draw (168,129.4) node [anchor=north west][inner sep=0.75pt]    {$\mathrm{Unif}_{\mathcal{S}}$};
\draw (86,68.4) node [anchor=north west][inner sep=0.75pt]    {$\mu _{1}$};
\draw (86,147.4) node [anchor=north west][inner sep=0.75pt]    {$\mu _{2}$};
\draw (124,205.4) node [anchor=north west][inner sep=0.75pt]    {$\mu _{12}$};
\draw (299,55.4) node [anchor=north west][inner sep=0.75pt]    {$\mathfrak{M}( \mu _{1})$};
\draw (487,57.4) node [anchor=north west][inner sep=0.75pt]  [color={rgb, 255:red, 0; green, 0; blue, 0 }  ,opacity=1 ]  {$T{_{\mathrm{Unif}_{\mathcal{S}}}^{\mu _{1}}}$};
\draw (508,123.4) node [anchor=north west][inner sep=0.75pt]    {$\mathrm{id}$};
\draw (437,180.4) node [anchor=north west][inner sep=0.75pt]  [color={rgb, 255:red, 0; green, 0; blue, 0 }  ,opacity=1 ]  {$T{_{\mu _{1}}^{\mu _{2}}}$};
\draw (286,180.4) node [anchor=north west][inner sep=0.75pt]    {$\mathfrak{M}^{-1}\left( T_{\mu _{1}}^{\mu _{2}}\right)$};
\draw (146,92.4) node [anchor=north west][inner sep=0.75pt]  [font=\footnotesize,color={rgb, 255:red, 0; green, 0; blue, 0 }  ,opacity=1 ]  {$T{_{\mathrm{Unif}_{\mathcal{S}}}^{\mu _{1}}}$};
\draw (83,107.4) node [anchor=north west][inner sep=0.75pt]  [font=\footnotesize,color={rgb, 255:red, 0; green, 0; blue, 0 }  ,opacity=1 ]  {$T{_{\mu _{1}}^{\mu _{2}}}$};
\draw (116,173.4) node [anchor=north west][inner sep=0.75pt]  [font=\footnotesize,color={rgb, 255:red, 0; green, 0; blue, 0 }  ,opacity=1 ]  {$T{_{\mathrm{Unif}_{\mathcal{S}}}^{\mu _{12}}}$};
\draw (140,251.4) node [anchor=north west][inner sep=0.75pt]    {$\mathcal{W}$};
\draw (484,255.4) node [anchor=north west][inner sep=0.75pt]    {$\mathcal{T}$};
\draw (505,184.4) node [anchor=north west][inner sep=0.75pt]  [font=\small,color={rgb, 255:red, 0; green, 0; blue, 0 }  ,opacity=1 ]  {$\left( T_{\mathrm{Unif}_{\mathcal{S}}}^{\mu _{1}}\right)^{-1}$};
\draw (162,198.4) node [anchor=north west][inner sep=0.75pt]  [font=\fontsize{0.44em}{0.53em}\selectfont,color={rgb, 255:red, 0; green, 0; blue, 0 }  ,opacity=1 ]  {$( T_{\mathrm{Unif}_{\mathcal{S}}}^{\mu _{1}})^{-1} \#\mathrm{Unif}_{\mathcal{S}}$};

\end{tikzpicture}
\caption{The relation between Wasserstein space and transport space.  The optimal transport from $\mu_{1}$ to $\mu_2$, i.e., $T_{\mu_1}^{\mu_2}=F_{\mu_2}^{-1}\circ F_{\mu_1}$, can be also regarded as the transport map from the uniform distribution  to the measure $\mu_{12}(A):=F_{\mu_2}^{-1}\circ F_{\mu_1}(A) $ and thus $\mathfrak{M}^{-1}(T_{\mu_1}^{\mu_2})=\mu_{12}. $} \label{fig:WT}
\end{figure}

Proposition \ref{prop:iso} implies  that the  transport space  is isometric to the Wasserstein space.  
The relationship between $\dwsp$ and $\mcalT$ is illustrated in Figure \ref{fig:WT}. The McCann interpolation \citep{mccann1997} reveals that $\dwsp$ is a 
uniquely geodesic space, where for any elements $x,y, \, x \ne y$  there exists a uniquely defined (constant speed) geodesic that 
connects $x$ and $y$; 
by Proposition \ref{prop:iso}, the transport space is  then also a uniquely geodesic space.
}

A scalar multiplication  operation in the transport space \citep{mull:21:3}, 
$$
\alpha \odot T(u):=\left\{\begin{array}{cc}
u+\alpha\{T(u)-u\}, & 0<\alpha\leq1 \\
u, & \alpha=0 \\
u+\alpha\left\{u-T^{-1}(u)\right\}, & -1 \leq \alpha<0
\end{array}\right.
$$
induces a geodesic on $\mcalT$ from $\mathrm{Unif}_{\wdomain}$ to $T$, denoted by $u\odot T$ for all $u\in[-1,1]$. Proposition S.3 in the Supplement shows that  there exists a  constant-speed geodesic between \(T^{-1} = (-1) \odot T\) and \(T\), implying that \(\mathcal{T}\) is symmetric with respect to the multiplication \(\odot\). This motivates the introduction of a binary relation \(\sim\) on \(\mathcal{T}\), defined as \(T_1 \sim T_2\) if and only if there exists \(a \in [0,1]\) such that \(T_1 = a \odot T_2\) or \(T_2 = a \odot T_1\).
\begin{prop}\label{prop-equclass}
	$\sim $ is an equivalence relation on $\mcalT$.
\end{prop} 

The equivalence class of $T\in\mcalT$ is denoted as $[T]_{\sim}$, and for each $T'\in [T]_{\sim}$, $T'$ resides on the extended geodesic $\mathrm{id}+u(T-\mathrm{id}) $.  One needs to fix  the norm of $T_{0}$ to ensure the identifiability of the proposed model. Motivated by   $\|T\|_{1}=\|T^{-1}\|_{1}$ for all $T\in\mathcal{T}$, which is easy to verify by Fubini's Theorem, we opt to use the metric $d_{W,1}$ to quantify the norm of $T$ within the transport space $\mathcal{T}$  abbreviated as  $d_{W}$.  
When $p\neq 1$, in general  $\|T\|_{p}=\|T^{-1}\|_{p}$ does not hold and two distinct values for $\|T\|_{p}$ and $\|T^{-1}\|_{p}$ need to be chosen.  The results presented in this paper can be extended to this general scenario, with minor but tedious modifications for which we do not give the details.  Since $[T_0]_{\sim}$ is an equivalence class, one can choose  $\mathrm{id}+u(T_{0}-\mathrm{id})$ for any $u>0$ as the representative of $[T_0]_{\sim}$.
\bco
\begin{figure}[tbp]
\centering
\tikzset{every picture/.style={line width=0.75pt}} 
\begin{tikzpicture}[x=0.75pt,y=0.75pt,yscale=-1,xscale=1]
\draw [color={rgb, 255:red, 74; green, 144; blue, 226 }  ,draw opacity=1 ]   (368.62,146.97) .. controls (415.46,138.97) and (460.61,192.72) .. (542.76,160.49) ;
\draw [shift={(544,160)}, rotate = 338.15] [color={rgb, 255:red, 74; green, 144; blue, 226 }  ,draw opacity=1 ][line width=0.75]      (0, 0) circle [x radius= 3.35, y radius= 3.35]   ;
\draw [shift={(368.62,146.97)}, rotate = 350.31] [color={rgb, 255:red, 74; green, 144; blue, 226 }  ,draw opacity=1 ][fill={rgb, 255:red, 74; green, 144; blue, 226 }  ,fill opacity=1 ][line width=0.75]      (0, 0) circle [x radius= 3.35, y radius= 3.35]   ;
\draw [color={rgb, 255:red, 74; green, 144; blue, 226 }  ,draw opacity=1 ] [dash pattern={on 0.84pt off 2.51pt}]  (544,160) .. controls (596.84,142.82) and (604,155) .. (626,144) ;
\draw [color={rgb, 255:red, 208; green, 2; blue, 27 }  ,draw opacity=1 ]   (217.88,129.12) .. controls (262.16,109.61) and (321.35,155.65) .. (368.62,146.97) ;
\draw [shift={(215.19,130.39)}, rotate = 333.01] [color={rgb, 255:red, 208; green, 2; blue, 27 }  ,draw opacity=1 ][line width=0.75]      (0, 0) circle [x radius= 3.35, y radius= 3.35]   ;
\draw [color={rgb, 255:red, 208; green, 2; blue, 27 }  ,draw opacity=1 ] [dash pattern={on 0.84pt off 2.51pt}]  (96.53,147.29) .. controls (134.35,145.75) and (175.56,153.94) .. (215.19,130.39) ;
\draw   (98,152) .. controls (98,156.67) and (100.33,159) .. (105,159) -- (223,159) .. controls (229.67,159) and (233,161.33) .. (233,166) .. controls (233,161.33) and (236.33,159) .. (243,159)(240,159) -- (361,159) .. controls (365.67,159) and (368,156.67) .. (368,152) ;
\draw   (623,141) .. controls (623,136.33) and (620.67,134) .. (616,134) -- (506,134) .. controls (499.33,134) and (496,131.67) .. (496,127) .. controls (496,131.67) and (492.67,134) .. (486,134)(489,134) -- (376,134) .. controls (371.33,134) and (369,136.33) .. (369,141) ;
\draw     ;
\draw    (368.62,146.97) ;
\draw [shift={(368.62,146.97)}, rotate = 0] [color={rgb, 255:red, 0; green, 0; blue, 0 }  ][fill={rgb, 255:red, 0; green, 0; blue, 0 }  ][line width=0.75]      (0, 0) circle [x radius= 3.35, y radius= 3.35]   ;
\draw (351,120) node [anchor=north west][inner sep=0.75pt]   [align=left] {id};
\draw (539,172.4) node [anchor=north west][inner sep=0.75pt]    {$T$};
\draw (204,104.4) node [anchor=north west][inner sep=0.75pt]    {$T^{-1}$};
\draw (209,175.4) node [anchor=north west][inner sep=0.75pt]    {$\left[ T^{-1}\right]_{\sim }$};
\draw (481,99.4) node [anchor=north west][inner sep=0.75pt]    {$[ T]_{\sim }$};
\end{tikzpicture}
\caption{An illustration of the equivalence class and geodesic.\label{fig:geod} }
\end{figure}
\fi

We quantify the overall direction of a transport as follows, 
\begin{equation}\label{eq:def-sign}
	\sign(T):= \operatorname{sign}\left(\int_{\wdomain}\{T(u)-u\} \diff u\right),
\end{equation}
where 
$\sign(T)=1$ represents the case where the overall direction of the mass transfer from $\mathrm{Unif}{\wdomain}$ to $T\# \mathrm{Unif}{\wdomain}$ predominantly is from left to right. Further discussions relating to the geometry  of the transport space can be found in the Supplement.
\suppcontent{
\begin{prop}\label{prop-sign}
	$\sign(\alpha\odot T)=\sign(\alpha)\sign(T) $ for all $\alpha\in[-1,1]$ and $T\in\mcalT$.
\end{prop}
}
\section{Modeling transport processes}\label{sec:meth}

\subsection {Transport model} \label{sec:basic}
We aim for an efficient representation of transport  processes $T(t)$,  $t \in \tdomain$ for a compact interval $\tdomain$, which are obtained by centering distributional processes as described above. 
 To address the challenge that Euclidean based methods such as 
 functional principal component analysis are not anymore applicable, due to the absence of inner products and linear  projections, we assume that  a realized transport process $T(t)$ possesses a characteristic  common transport pattern for all $t\in\tdomain$, where this pattern is specific for each realization. It corresponds to an overall random transport that characterizes the specific realization of the process. In functional data analysis this feature  is captured by functional principal components that correspond to trajectory-specific random effects. 

 More specifically, in analogy to the decomposition of Euclidean-valued functional data into a mean function and a stochastic part,  we assume  that the centered transport processes $T(t)$ can be decomposed into a scalar random function $U(t)$ that serves as a scalar multiplier in the transport space and a characteristic overall transport $T_0$, 
\begin{equation}\label{tm1}
	T(t)= U  (t)\odot T_{0}, \text{ for all }t\in\tdomain,
\end{equation}
where $T_0$ is a random element in $\mcalT$ associated with  each realization of the transport process.  The scalar multiplier function is  a stochastic process that takes values in {$(-1,1)$}  and  is derived from an  unconstrained process $Z$ through a transformation $g$ as follows,  
\be \quad U(t)=g(Z(t)), \,\, Z(t) \in \reall, \,\, \E [Z(t)]=0, \,\,   g:\reall\mapsto (-1,1), \,\, g \text{ is bijective,  for all } t \in \tdomain.\label{tm2} \ee
The mean zero stochastic process $Z(t)$ in conjunction with the  bijective map $g:\reall\mapsto (-1,1)$ further characterizes the transport process $T$, where $T(t)$ resides in $ \{T: T\in  [T_{0}]_{\sim}\}\cup\{T:T\in [T_0^{-1}]_{\sim}\}$, which includes the geodesic from $T_{0}^{-1}$ to $T_{0}$.  

For some situations it is appropriate and  advantageous to further assume   that the process $Z$ is a Gaussian process, a property that can be  harnessed to obtain methods for the important case where the distribution-valued trajectories are only observed on a discrete grid of time points that might be sparse. 
In Section \ref{sec:appli}, we show that the transport process model, as defined by equations \eqref{tm1} and \eqref{tm2}, is well-suited for practical applications, while  the assumptions it entails are not overly restrictive. Proposition S.1  in the Supplement   demonstrates  
that the stochastic transport process  \eqref{tm1} is well-defined.

Throughout we assume that one has a sample $\{T_{i}(t)\}_{i=1}^n$ of i.i.d. realizations  of the transport process $T(t)$ that permits the decomposition in \eqref{tm1}, \eqref{tm2} and furthermore that the norms $\|T_{i0}\|_{1}$ are the same for all $i=1,\dots,n$. To ensure the identifiability of the proposed model 
in \eqref{tm1}, \eqref{tm2}, 
it turns out to be necessary to preselect the norm of $T_{0}$. As mentioned,  it is often not possible to observe the full process $T_{i}(t)$ for all $t\in\tdomain$ and measurements may be available only at a few  discrete time points $\{t_{ij}\}_{j=1}^{N{i}}$ for the $i$th subject. An additional difficulty is that  in distributional data analysis 
the distributions serving as data atoms frequently are unknown and only random samples generated by these distributions are available. In this situation, a standard  pre-processing step is to estimate the underlying distributions first and to work with estimated transports $\hat{T}_{ij}$. Further discussion of this issue can be found in the Supplement.


Aiming  to represent and recover the transport trajectories $T_{i}(t)$ for all $t\in\tdomain$ based on the available discrete observations $\{(t_{ij},\hat{T}_{ij})\}_{j=1}^{N_i}$, we first require  a reliable estimate of the baseline transport $T_{i0}$ for each subject $i$ in the framework of model \eqref{tm1}.  Since $\alpha\odot T_{i0}$ belongs to different equivalence classes for positive and negative $\alpha$, it is necessary to estimate $T_{i0}$ and its inverse $T_{i0}^{-1}$ separately. We define $\hat{I}_{i}^{+}=\{j:\sign(\hat{T}_{ij})>0\}$ and $\hat{I}_{i}^{-}=\{j:\sign(\hat{T}_{ij})<0\}$ as the index sets for positive and negative  $\{\hat{T}_{ij}\}_{j=1}^{N_{i}}$, respectively.  Denoting by $\tilde{T}_{i0}^{+}$ and $\tilde{T}_{i0}^{-}$  the Fr\'echet integrals  \citep{petersen2016frechet} with respect to $\hat{I}_{i}^{+} $ and $\hat{I}_{i}^{-} $, 
$$\tilde{T}_{i0}^{+}=\argmin{T\in\mcalT}\frac{1}{|\hat{\idomain}_{i}^{+} |}\sum_{j\in \hat{I}_{i}^{+}}\int_{\mathcal{S}} \{\hat{T}_{ij}(u)-T(u)\}^2\diff u,\,\,\tilde{T}_{i0}^{-}=\argmin{T\in\mcalT}\frac{1}{|\hat{\idomain}_{i}^{-} |}\sum_{j\in \hat{\idomain}_{i}^{-}}\int_{\mathcal{S}} \{\hat{T}_{ij}(u)-T(u)\}^2\diff u, $$
the solutions to these optimization problems are simply 
\begin{equation} \label{TT} \tilde{T}_{i0}^{+}(u)=\frac{1}{|\hat{\idomain}_{i}^{+} |}\sum_{j\in \hat{\idomain}_{i}^{+}}\hat{T}_{ij}(u)\text{   and   } \tilde{T}_{i0}^{-}(u)=\frac{1}{|\hat{\idomain}_{i}^{-} |}\sum_{j\in \hat{\idomain}_{i}^{-}}\hat{T}_{ij}(u).\end{equation}

We assume $\sign(T_{i0})>0$ for all $i=1,\ldots,n$ without loss of generality. Otherwise, the signs of $U_i(t)$ and $T_{i0}$ are not identifiable due to Proposition S.4 of the Supplement. Note that $\tilde{T}_{i0}^{+}$ and $\tilde{T}_{i0}^{-1}$ are estimators of representatives of equivalence classes $[T_{i0}]_{\sim}$ and $[T_{i0}^{-1}]_{\sim}$, and one can rescale $\tilde{T}_{i0}^{+}$ and $\tilde{T}_{i0}^{-}$ for any  $\kappa>0$ by
\begin{equation}\label{eq:def-kap}
	\hat{T}_{i\kappa}^{+}(u)=u+\frac{\kappa } {\|\tilde{T}_{i0}^{+}\|_{1}}   \{\tilde{T}_{i0}^{+}(u)-u\}\quad \text{ and }\quad \hat{T}_{i\kappa}^{-}(u)=u+\frac{\kappa} {\|\tilde{T}_{i0}^{-}\|_{1}}   \{\tilde{T}_{i0}^{-}(u)-u\}.
\end{equation}
Since  $\|\alpha\odot T_{i0}\|_{1}=|\alpha |\|T_{i0}\|_{1}$ for all $\alpha\in[-1,1]$,  $T_{i0}$ and $U(t)$ are not identifiable unless either $U(t)$ or the norm of $T_{i0}$   are specified.  
As mentioned before, $T_{i0}$ merely serves  as the representative of the   equivalence class $[T_{i0}]_{\sim}$ and one can choose any other representative $T_{i0}'=\mathrm{id}+\kappa(T_{i0}-\mathrm{id})$;  
this means we are free to  fix  the norm of $T_{i0}$ at a pre-specified value. 
This makes it possible to estimate the covariance functions for $U$ and $Z$, 
\begin{equation}\label{CD}
C(s,t)=\E[U(s)U(t)], \quad\quad D(s,t)=\E[Z(s)Z(t)].\end{equation}

If $T_{i}(t)$ is observed for all $t\in\tdomain $ without measurement errors,  $U$ and $Z$ processes can be represented by {$${U}_{i}(t)=\|T_{i}(t)\|_{1}\sign(T_{i}(t))/\|T_{i0}\|_{1}\text{ and }{Z}_{i}(t)=g^{-1}(\|T_{i}(t)\|_{1}\sign(T_{i}(t))/\|T_{i0}\|_{1}),$$ 
since $g$ is a bijective map.}
If measurements are only available at discrete time points $\{t_{ij}\}_{j=1}^{N_{i}}$ for each subject $i$, we use 
\begin{equation}\label{eq:def-hatUZ}
	\hat{U}_{i}(t_{ij})=\|\hat{T}_{ij}\|_1\sign(\hat T_{ij})/\|T_{i0}\|_1 \text{ and }\hat{Z}_{i}(t_{ij}) =g^{-1}(\|\hat{T}_{ij}\|_1\sign(\hat T_{ij})/\|T_{i0}\|_1)
\end{equation}
as estimators for $U_{i}(t_{ij})$ and $Z_{i}(t_{ij})$. Then $\hat{C}_{ijl}=\hat{U}_{i}(t_{ij})\hat{U}_{i}(t_{il}) $ and $\hat{D}_{ijl}=\hat{Z}_{i}(t_{ij})\hat{Z}_{i}(t_{il}) $ are the raw covariances for $C$ and $D$, respectively. To smooth the raw covariance, we adopt local linear smoothing,   in analogy to the approach in classical functional data analysis  \citep{yao2005functional,li2010uniform,zhang2016sparse}. For each $s,t\in\tdomain$, by taking $\mathrm{Raw}_{ijl}= \hat{C}_{ijl}$ or $=\hat{D}_{ijl}$ in equation \eqref{eq:def-ll-cov}, we  use $\hat{\beta}_{0}$ as  estimator for $C(s,t), D(s,t)$, respectively, with bandwidths $h$, a kernel $\K$ that is a symmetric density function on $[-1,1]$ and 
 \begin{equation}\label{eq:def-ll-cov}
	\begin{aligned}
		(\hat{\beta}_{0},\hat{\beta}_{1},\hat{\beta}_{2})
		=&\argmin{\beta_{0},\beta_{1},\beta_{2}}\sum_{i=1}^{n}w_{i}\sum_{j\neq l}\left\{ \mathrm{Raw}_{ijl} -\beta_{0}-
		\beta_{1}(t_{ij}-s)-\beta_{2}(t_{il}-t )\right\}^2\\
		&\quad\quad\quad\quad \times\K_{h}(t_{ij}-s)\K_{h}(t_{il}-t),
	\end{aligned}
\end{equation}
 with   $w_{i}=\{nN_{i}(N_{i}-1)\}^{-1}$ and  $\K_{h}(\cdot)=h^{-1}\K(\cdot/h)$. 
 
\subsection{Estimators for densely observed transport processes}\label{sec:meth-den}

We first consider the dense case where $\underline{N}:=\min\{N_i\}_{i=1}^{n}\rightarrow\infty$.  
In this case, knowledge of $\|T_{i0}\|_{1}$ is not required in order to  obtain a consistent estimator of $T_{i}(t)$ since 
\begin{equation}\label{eq:T-rescale}
	T_{i}(t)=U_{i}(t)\odot T_{i0}(u)=(U_{i}(t)/\kappa) \odot (u+\kappa/\|T_{i0}\|_{1}\{T_{i0}(u)-u\} )  \text{ for all }\kappa>0.
\end{equation}
This means that 
we can define a rescaled version of $C(s,t)$ with $C_{\kappa}(s,t)=\|T_{i0}\|_{1}^2\kappa^{-2} C(s,t)$, which is the covariance function of $ U_{i,\kappa}(t)= \|T_{i0}\|_1 U  _{i}(t)/ \kappa$. Assume $C(s,t)$ admits the eigendecomposition 
\be \label{KLC} C(s,t)=\E[U(s)U(t)]=\sum_{k=1}^{\infty}\lambda_{k}\phi_{k}(s)\phi_{k}(t), \quad  k =1,\dots, \infty,\ee with an orthonormal system of eigenfunctions $\phi_k$, eigenpairs $\{(\lambda_{k},\phi_{k})\}_{k=1}^{\infty}$ and positive eigengaps  $\lambda_k-\lambda_{k+1}>0$ for all $k$ for the linear auto-covariance operator of $U$. 
Since $C_{\kappa}(s,t)$ is proportional to $C(s,t)$,  the eigenfunctions of $C_{\kappa}(s,t)$ and $C(s,t)$ are identical and the eigenvalues of $C_{\kappa}(s,t)$ are proportional to $\lambda_{k}$. The process $ U  _{i}(t)$ and its corresponding rescaled version $ U  _{i,\kappa}(t) $ admit the Karhunen-Lo\`eve expansion
\be \label{Urep}  U  _{i}(t)=\sum_{k=1}^{\infty}\xi_{ik}\phi_{k}(t) \text{ and }  U  _{i,\kappa}(t)=\sum_{k=1}^{\infty}\xi_{ik,\kappa}\phi_{k}(t),\ee
where $\xi_{ik}=\int U  _{i}(t)\phi_{k}(t)\diff t$ and $\xi_{ik,\kappa}=\|T_{i0}\|_{1}\xi_{ik}/\kappa $.

As previously mentioned, $\hat{C}_{ijl,\kappa}:=\|\hat{T}_{ij}\|_{1}\|\hat{T}_{il}\|_{1}\sign(\hat{T}_{ij})\sign(\hat{T}_{il})/\kappa^2 $ can be used as  raw covariance for $C_{\kappa}(t_{ij},t_{il})$. Specifically in this subsection, we further assume that the $t_{ij}$ are random samples from $\mathrm{Unif}(0,1)$ without loss of generality; the uniform distribution assumption can be easily relaxed but at the cost of more involved notation. We replace $\mathrm{Raw}_{ijl}$  with $\hat{C}_{ijl,\kappa}$ in equation \eqref{eq:def-ll-cov} to obtain the  covariance estimator $\hat{C}_{\kappa}(s,t)=\hat{\beta}_{0}$. The estimated covariance function $\hat{C}_{\kappa}(s,t)$ admits an empirical eigendecomposition 
$$\hat{C}_{\kappa}(s,t)=\sum_{k=1}^{\infty}\hat{\lambda}_{k,\kappa}\hat{\phi}_{k}(s)\hat{\phi}_{k}(t),$$
where $\hat{\lambda}_{k,\kappa}$ and  $\hat{\phi}_{k}$ are estimators for ${\lambda}_{k,\kappa}=\|T_{i0}\|\lambda_{k}/\kappa$ and  ${\phi_{k}}$, respectively.
Using the  $\hat{\phi}_{k}$, we can recover $ U  _{i,\kappa}(t)$ in \eqref{Urep} with \begin{equation} \label{Uhat} \hat{ U  }_{i,\kappa}(t)=\sum_{k=1}^{J_n}\hat{\xi}_{ik,\kappa}\hat{\phi}_{k}(t), \quad \hat{\xi}_{ik,\kappa}=\frac{1}{N_i}\sum_{j=1}^{N_i}\frac{\|\hat{T}_{ij}\|\sign(\hat{T}_{ij})}{\kappa}\hat\phi_{k}(t_{ij}),\end{equation}
 where we will consider $J_n\rightarrow\infty$ in the theory. 

The proposed  final  estimator for ${T}_{i}(t)$ is obtained by combining \eqref{TT},   \eqref{eq:def-kap} and \eqref{Uhat}, 
\begin{equation}\label{eq:def-estTi}
	\hat{T}_{i}(t)=\hat{ U  }_{i,\kappa}(t)\odot \hat{T}_{i\kappa}^{+}(u).
\end{equation} 
Here we assume the index set $\hat{I}_{i}^{+}$ is non-empty without loss of generality. In practice, if all the $\sign(\hat T_{ij})$ are negative for a specific $i$, it is expedient to use $\hat{T}_{i}(t)=-\hat{ U  }_{i,\kappa}(t)\odot \hat{T}_{i\kappa}^{-}(u)$ instead, as this does not affect the asymptotic behavior of $\hat{T}_{i}(t)$. 

\subsection{Latent Gaussian processes  for sparsely observed data}\label{sec:meth-spa}
When ${N}_{i}$ is finite, the estimator \eqref{eq:def-estTi} is not consistent due to  approximation bias. In analogy to the approach of  \cite{yao2005functional}, this problem can be overcome by further assuming that $Z$ is a Gaussian process. This makes it possible to evaluate  the conditional expectation of $T_i(t)$ given the data through the best predictor, which under Gaussianity is the best linear predictor of $T_{i}(t)$ for which one has an explicit form. 

In contrast to the dense case, where knowledge of $\|T_{i0}\|_1$ is not required for predicting $T_{i}(t)$, an estimator for $\|T_{i0}\|_1$ is needed in the sparse case due to the nonlinearity of $g$. To this end, we pre-fix $\|T_{i0}\|_{1}$. Recalling  that $\hat{Z}_{ij}=g^{-1}(\|\hat{T}_{ij}\|_1\sign(\hat{T}_{ij})/\|T_{i0}\|_{1})$ and  replacing the raw covariance $\mathrm{Raw}_{ijl}$ by $\hat{D}_{ijl}=\hat{Z}_{ij}\hat{Z}_{il}$  in equation \eqref{eq:def-ll-cov}, we  obtain the local linear estimator $\hat{D}(s,t)$ for the covariance function $D(s,t)$.

Assume $D(s,t)$ and $\hat{D}(s,t) $ admit  eigendecompositions $${D}(s,t)=\sum_{l=1}^{\infty}\eta_{l}\psi_{l}(s) \psi_{l}(t)\text{ and }  \hat{D}(s,t)=\sum_{l=1}^{\infty}\hat\eta_{l}\hat\psi_{l}(s) \hat\psi_{l}(t),$$ where $\{\psi_{l}\}_{l=1}^{\infty}$, $\{\hat\psi_{l}\}_{l=1}^{\infty}$ are  eigenfunctions and $\eta_{l} $, $\hat\eta_{l} $ are the paired eigenvalues, so that the corresponding Karhunen-Lo\`eve expansions for $Z_{i}(t)$ and $\hat{Z}_{i}(t)$ are ${Z}_{i}=\sum_{l=1}^{\infty}\chi_{il}\psi_{l} $ and $\hat{Z}_{i}=\sum_{l=1}^{\infty}\hat\chi_{il}\hat\psi_{l}$  with functional principal component scores   $\chi_{il}=\int Z_{i} \psi_{l} $ and $\hat{\chi}_{il}=\int \hat{Z}_{i} \hat{\psi}_{l}$. 

With
 $\bm{\hat{Z}}_{i}=(\hat{Z}_{i1},\ldots,\hat{Z}_{iN_{i}})^{T}$, $\bm{{Z}}_{i}=({Z}_{i}(t_{i1}),\ldots,{Z}_{i}(t_{iN_i})  )^{T}$; $\hat{\Psi}_{il}=(\hat{\psi}_{l}(t_{i1}),\ldots,\hat{\psi}_{l}(t_{iN_i})  )^{T}$ and ${\Psi}_{il}=({\psi}_{l}(t_{i1}),\ldots,{\psi}_{l}(t_{iN_i}) )^{T}$; $[\hat{\Sigma}_{i}]_{lj}=\hat{D}(t_{il},t_{ij})$ and $[{\Sigma}_{i}]_{lj}={D}(t_{il},t_{ij})$,  $\Sigma_{i} $ is positive definite if the observations $\{t_{ij}\}_{j=1}^{N_{i}}$ are distinct for each subject; formally:
\begin{lemma}\label{lem-pace:Sig}
	Assume  $\{\psi_{j}\}_{j=1}^{\infty}$ are uniformly bounded  in $j$. If the $\{t_{ij}\}_{j=1}^{N_{i}}$ are distinct, then $\Sigma_{{i}}$ is positive definite and thus invertible. 
\end{lemma}\label{thm-pace:Sig}
As a consequence, if $Z_i$ is a Gaussian process, the best linear predictor of $\chi_{il} $ given $\bm{Z}_{i}$ is  \be \label{bp} \tilde{\chi}_{il}={\eta}_{l} {\Psi}_{il}^{T}{\Sigma}^{-1}_{i}\bm{Z}_{i}.\ee
Combining Lemma \ref{lem-pace:Sig} with Lemma A.3 in \cite{facer2003nonparametric}, one obtains  that $\hat{\Sigma}^{-1}_{i} $ is also invertible for large sample sizes $n$. 
Using   $\hat{\chi}_{il}=\hat{\eta}_{l} \hat{\Psi}_{il}^{T}\hat{\Sigma}^{-1}_{i}\bm{\hat{Z}}_{i} $ as  estimator of $\tilde{\chi}_{il} $ and  $\hat{Z}_{i}^{J}(t)=\sum_{l=1}^{J}\hat{\chi}_{il}\hat\psi_{l}(t) $ as  estimator for ${Z}_{i}(t)$, we arrive at the following predictor for   $T_{i}$,
\begin{equation}\label{eq:def-T-est-S}
	\hat{T}_{i}^{J}(t)=g(\hat{Z}_{i}^{J}(t))\odot \hat{T}_{i}^{+} \text{ with } \hat{T}_{i}^{+}=u+\frac{\|T_{i0}\|_1}{\|\tilde{T}_{i0}^{+}\|_1}    \{\tilde{T}_{i0}^{+}(u)-u\},
\end{equation}
where $J$ is a fixed positive integer.

\suppcontent{
\subsection{From distribution-generated data to estimated transports  $\hat{T}_{ij}$}\label{sec:hatT}
As already mentioned,  the distributions $T_{ij}$ are often unknown and only random samples generated by these distributions are available for further analysis, i.e., the available data are $
\{(t_{ij},{x}_{ijk}) \}_{k=1}^{m_{ij}},$ for  $i=1,\ldots,n$ and $ j=1,\ldots, N_{i} $. Here $\{{x}_{ijk} \}_{k=1}^{m_{ij}} $ are random samples drawn from the probability measures corresponding to $T_{ij}$. Specifically, in the case where $X_{i}(t)$ is a distribution process and  $T_{ij}$ is the optimal transport from $\mu_{\oplus}(t_{ij})$ to $X_{i}(t_{ij})$, where $\mu_{\oplus}(\cdot )$ is the Fr\'echet mean of $X_{i}(\cdot)$, the true  observations $\{x_{ijk} \}_{k=1}^{m_{ij}}$ are the random samples from each $X_{i}(t_{ij})$. Based on $\{x_{ijk} \}_{k=1}^{m_{ij}}$, consistent estimates of cumulative distribution functions $F_{i}(t_{ij})$ and quantile functions $F^{-1}_{i}(t_{ij})$ of $X_{i}(t_{ij}) $, denoted by $\hat{F}_{ij}$ and $\hat{F}_{ij}^{-1}$, are readily available \citep{falk1983relative, leblanc2012estimating, petersen2016functional}. 

For fixed designs, where the $\{t_{ij}\}_{j=1}^{N_{i}}$ differ across $j$ but are the same across $i=1,\ldots,n$, the quantile function of the Fr\'echet mean $\mu_{\oplus}$ at $t_{ij}$ is estimated by  $\hat{F}_{\oplus}^{-1}(t_{ij})=\sum_{i=1}^{n}\hat{F}^{-1}_{ij}/n$. In the case of random design, where $\{t_{ij}\}_{j=1}^{N_{i}}$ are random samples from a probability measure on the domain $\tdomain$, one may employ  local Fr\'echet regression to obtain  the Fr\'echet means $\hat{\mu}_{\oplus}$, 
$$\hat{\mu}_{\oplus}(t)=\argmin{p\in\wsp} \frac{1}{n}\sum_{i=1}^{n}\frac{1}{N_{i}}\sum_{j=1}^{N_i}\hat\omega(t_{ij},t,h)d^{2}(\hat{F}_{ij},p) .$$
Here, $\hat\omega(s,t,h)=\hat{\sigma}_{0}^{-2}\mathrm{K}_{h}(s-t)\{\hat{\kappa}_{2}-\hat\kappa_{1}(s-t) \} $, $\hat{\kappa}_{r}=n^{-1}\sum_{i=1}^{n}N_{i}^{-1}\sum_{j=1}^{N_i}\mathrm{K}_{h}(t_{ij}-t)(t_{ij}-t)^r$ for $r=0,1,2$ and $\hat{\sigma}_{0}^2=\hat{\kappa}_{0}\hat{\kappa}_{2}-\hat{\kappa}_{1}^2$. Having $\hat{F}_{ij}$ and $\hat{\mu}_{\oplus}$ in hand, one can obtain optimal transport estimates $\hat{T}_{ij}=\hat{F}^{-1}_{ij}\circ \hat{F}_{\oplus}(t_{ij})$, where $\hat{F}_{\oplus}$ is the cumulative  distribution function of  $\hat{\mu}_{\oplus}$.

For the asymptotic analysis in Section \ref{sec:theo-T} we will require that  $m=\min_{i,j}m_{ij}$ satisfies $m=m(n)  \rightarrow \infty$ as 
$n  \rightarrow \infty.$ We will demonstrate that if $m$ increases rapidly enough relative to the sample size $n$, the effect of estimating the distributions from the data they generate is asymptotically negligible.  This will be based on a result of the type  $\sup_{i,j}\E d_{W}(\hat{T}_{ij},T_{ij}) =o_{P}(\tau_{m})$ for a suitable null sequence $\tau_m$.

}
\section{Theoretical results} \label{sec:theo}

\subsection{General assumptions}\label{sec:theo-assum}
The following mild assumptions  are needed for the theory. 
\begin{assum}\label{asm:Ti0}
	There exists a constant $c>0$ such that $\E|[ \int \{T_{0}(u)-u\}\diff u ]^{-1}|\leq c$. 
\end{assum}
\begin{assum}\label{asm:iid}
	{The times $t_{ij}$ where processes are observed are distributed on the interval $\tdomain$ according to a distribution which has a continuous density that is bounded below away from 0. The times $t_{ij}$,   processes   $Z(t)$ and characteristic transports $T_{0}$ are jointly independent.}
\end{assum}
\begin{assum}\label{asm:generalz}
	The stochastic process  $Z$ satisfies  $\sup_{t\in[0,1]}\prob(|Z(t)|\leq x)\leq cx $ for a  $c>0$.
\end{assum}
\begin{assum}\label{asm:gmap}
	The bijective map $g$ is symmetric and convex on $(-\infty,0]$. Moreover, for all $\varrho>0$, $g^{-1}$ is Lipschitz continuous on $[-1+\varrho,1-\varrho]$, that is, there exists a constant $L_{\varrho}$ such that $\frac{g^{-1}(x_{1})-g^{-1}(x_{2})}{x_{1}-x_{2}}\leq L_{\varrho}.$
\end{assum}

When the integral $\int \{T_{i0}(u)-u\}\diff u$ is close to $0$, it is more likely that the signs of $\hat{T}_{ij}$ and ${T}_{ij}$ differ, even while  $d_W(\hat{T}_{ij},T_{ij})$ is small. Assumption \ref{asm:Ti0} requires that the integral $\int \{T_{i0}(u)-u\}\diff u$ is not close to 0 and is needed  to establish  the consistency of $\sign(\hat{T}_{ij})$. Similar assumptions have been adopted for distributional time series models \citep{mull:21:3}. Assumption  \ref{asm:iid}  requiring  
the independence of  functional trajectories and observed time points is a standard assumption  in functional data analysis \citep{yao2005functional, zhang2016sparse, zhou2022theory} and also includes  the characteristic transport $T_{0}$.  Assumption \ref{asm:generalz} is needed to show that the sign of the estimated transport is consistent with its true version; specifically it is satisfied if  $Z$ is a Gaussian process, where  $\prob(0\leq Z(t)\leq x |t)\leq cx$ with $c=\sup_{t\in \tdomain}{2\pi \E[Z^2(t)]}^{-1/2}$. Note that for  bijective maps from $(-1,1)$ to $\reall$, Lipschitz continuity can only be satisfied on a compact subset of $(-1,1)$.  Assumption \ref{asm:gmap} 
is needed for the analysis of the asymptotic behavior of the process $Z$. Examples of maps that satisfy Assumption 4 and are of practical interest can be found in Supplement.
\suppcontent{
Some examples of maps that satisfy Assumption 4 and are of practical interest: 
	\begin{itemize}
		\item[1.] $g_1(x)=\frac{2}{\pi}\arctan(x)$. Then $g_1^{-1}(x)=\tan(\pi x/2)$ and for all $x_{1},x_{2}\in [-1+\varrho,1-\varrho]$
		\begin{align*}
			\sup_{x_{1},x_{2}\in[-1+\varrho,1-\varrho] }\frac{g_1^{-1}(x_{1})-g_1^{-1}(x_{2})}{x_{1}-x_{2}}\leq\sup_{x\in[-1+\varrho,1-\varrho] } \frac{\diff \tan(\pi x/2)}{\diff x}\leq\frac{\pi}{1-\cos(\pi \varrho)} .
		\end{align*}
		\item[2.] $g_2(x)=\frac{\sqrt{1+4x^2}-1}{2x}$. Then $g_2^{-1}(x)=\frac{x}{1-x^2}$ and for all $x_{1},x_{2}\in [-1+\varrho,1-\varrho]$,
		\begin{align*}
			\sup_{x_{1},x_{2}\in[-1+\varrho,1-\varrho]}\frac{g_2^{-1}(x_{1})-g_2^{-1}(x_{2})}{x_{1}-x_{2}}\leq \sup_{x\in[-1+\varrho,1-\varrho] } \frac{\diff \left(\frac{x}{1-x^2} \right)}{\diff x}\leq \frac{(\varrho-1)^2+1}{\varrho^2(\varrho^2-2)^2}  .
		\end{align*}
		\item[3.] $g_3(x)=\frac{e^x-1}{e^x+1}$. Then  $g_3^{-1}(x)=\log \frac{1+x}{1-x}$ and for all $x_{1},x_{2}\in[-1+\varrho,1-\varrho]$,
		\begin{align*}
			\sup_{x_{1},x_{2}\in[-1+\varrho,1-\varrho]}\frac{g_3^{-1}(x_{1})-g_3^{-1}(x_{2})}{x_{1}-x_{2}}\leq \sup_{x\in[-1+\varrho,1-\varrho] } \frac{\diff \left(\log \frac{1+x}{1-x} \right)}{\diff x}\leq \frac{2}{\varrho(2-\varrho)}  .
		\end{align*}
	\end{itemize}
	}
\begin{assum}\label{asm:kernel}
	The kernel $\K$ is a bounded continuous symmetric probability density function on $[-1,1]$ satisfying 
	$\int u^{2}\K (u)\diff u <\infty,\  \int \K^{2}(u)\diff u<\infty.$
\end{assum}
\begin{assum}\label{asm:cov} \quad
	(a).  The covariance function  $C(s,t)$ in \eqref{CD} has bounded second order derivatives and its corresponding eigenfunctions $\{\phi_{j}\}_{j=1}^{\infty}$ are uniformly bounded  in $j$.
	
	  \no (b).   The covariance function $D(s,t)$ in \eqref{CD}  has bounded second order derivatives and its corresponding eigenfunctions $\{\psi_{j}\}_{j=1}^{\infty}$ are uniformly bounded  in $j$.

\end{assum}	

These assumptions 
are common and widely adopted in kernel smoothing and functional data analysis \citep{yao2005functional, zhang2016sparse};   Assumption \ref{asm:cov} is used to derive the consistency for the covariance of  $U$ and $Z$. Since in general the process $Z$ is unbounded, 
one needs to consider an increasing sequence $\rho=\rho(n)$ and correspondingly increasing  Lipschitz constants  $L_{\varrho}$ in Assumption \ref{asm:gmap},  
 in dependence on the increasing sequence 
\begin{equation}\label{eq:Mn}
	M_{n}:=\max_{i=1,\ldots,n}\sup_{t\in\tdomain}|Z_{i}(t)|,
\end{equation}
in order to obtain  asymptotic convergence  for $\hat{D}(s,t)$; by Theorem 5.2 in \cite{adler1990introduction}, if $Z$ is Gaussian, then $M_{n}$ is polynomial in $\log n$. 

\subsection{Asymptotics}\label{sec:theo-T}

Note that the estimated $\hat{T}_{ij}$ might have a sign that differs from that of   $T_{ij}$, which can cause convergence problems as the  proposed estimators 
utilize  $\sign(\hat{T}_{ij})$ as per \eqref{TT}.  Therefore we need to quantify the probability of  the event  $\{\sign(\hat{T}_{ij})\neq \sign(T_{ij})\}$. For the case where  one does not observed the actual distributions $X(t_{ij})$ but instead only has $m_{ij}$ data that are generated by the distribution $X(t_{ij})$  we require  \begin{equation}  \label {m} m:=\min_{i,j}m_{ij}, \quad m=m(n)  \rightarrow \infty
\,\, \text{as} \,\, n  \rightarrow \infty, \end{equation}  i.e., that there is  a universal  lower bound $m$ for the number of observations
available for each distribution. We quantify the discrepancy between the actual and estimated distributions by   a sequence $\tau_{m}$ such that
\be \label{tau} \sup_{i,j}\E d_{W}(\hat{T}_{ij},T_{ij}) =o_{P}(\tau_{m}) .\ee
This leads to a corresponding bound on  the probability of the event $\{\sign(\hat{T}_{ij})\neq \sign(T_{ij})\}$. 
\begin{lemma}\label{lemma-sign}
	Assume $\{T_i(t)\}_{i=1}^{n}$ are generated from model \eqref{tm1} and \eqref{tm2} and   Assumptions \ref{asm:Ti0}  - \ref{asm:gmap} are satisfied. Then if  \eqref{tau} holds, 	
	$$\sup_{i=1,\ldots,n \atop j=1,\ldots,N_i}\prob\{\sign(\hat{T}_{ij})\neq \sign({T}_{ij} ) \}=O  (\tau_{m}) .$$
\end{lemma}

Note that $\tilde{T}_{i0}^{+}$ and $\tilde{T}_{i0}^{-1}$ are estimators of representatives of equivalence classes $[T_{i0}]_{\sim}$ and $[T_{i0}^{-1}]_{\sim}$, respectively. To quantify the discrepancy between $\tilde{T}_{i0}^{+}$ and $[T_{i0}]_{\sim}$, we define the distance between an equivalence class $[T]_{\sim}$ and a transport map $T'\in\mcalT$ as $d_{\sim}(T^{'};{[T]_{\sim}})=\inf_{p\in[T]_{\sim} } d_W (T',p)$. 
The following result  provides the consistency of $\tilde{T}_{i0}^{+}$ and $\tilde{T}_{i0}^{-}$ in terms of the distance $d_{\sim}$.
\begin{theorem}\label{thm-equclass}
	For $\tilde{T}_{i0}^{+}$ and $\tilde{T}_{i0}^{-}$ as defined in \eqref{eq:def-kap}, under Assumptions \ref{asm:Ti0} - \ref{asm:gmap}, 	$$d_\sim(\tilde{T}_{i0}^{+};{[T_{i0}]_{\sim}})=O_{P}(\tau_m)\text{ and }d_\sim(\tilde{T}_{i0}^{-};{[T_{i0}^{-1}]_{\sim}})=O_{P}(\tau_m),\quad \text{
	uniformly in $i$.}$$
\end{theorem}

If $T_{ij}$ is the optimal transport from $\mu_{\oplus}(t_{ij})$ to $X_{i}(t_{ij})$, where $\mu_{\oplus}(\cdot )$ is the Fr\'echet mean of $X_{i}(\cdot)$, one can directly obtain the convergence rate of $ d_W \{\hat{X}_{i}(t_{ij}),X_{i}(t_{ij}) \}=O_{P}(m_{ij}^{-1/4})$ \citep{panaretos2016amplitude} under suitable assumptions or alternatively  under different assumptions $d_W \{\hat{X}_{i}(t_{ij}),X_{i}(t_{ij}) \}=O_{P}(m_{ij}^{-1/3})$ 
on the set of absolutely continuous measures \citep{petersen2016functional}. Then the rate $\tau_m$ in  \eqref{tau} is $d_W (\hat{T}_{ij},T_{ij})=\max\{d_W(\hat{\mu}_{\oplus}(t_{ij}),\mu_{\oplus}(t_{ij})), \,  d_W(\hat{X}_{ij},X_{ij} ) \}$ \citep{mull:21:3}.

As a consequence, Corollary S.1 in the Supplement delineates  the convergence rate of the covariance functions $C$ and $D$ in \eqref{CD}, using  Lemma \ref{lem-pace:Sig} and arguments provided in \cite{zhang2016sparse}.   In the following, we use the average of the numbers of measurements $N_i$ that one has for each realization of the distributional  process, 
\be \label{Nbar} \bar{N}=n^{-1}\sum_{i=1}^{n}N_{i}. \ee

\suppcontent{
\begin{corollary}\label{thm-cov}
	Under Assumptions  \ref{asm:Ti0} -  \ref{asm:kernel}, for  $\bar{N}$ as in \eqref{Nbar}, 
	\begin{itemize}
		\item[1.] If  Assumption \ref{asm:cov}(a) holds, then
		$$\|\hat{C}-C\|=O_{P}\left(\frac{1}{\sqrt{n}}\left(1+\frac{1}{\bar Nh} \right)+h^{2}+\tau_m \right);$$  $$\|\hat{C}-C\|_{\infty}=O_{P}\left(\frac{\log n}{\sqrt{n}}\left(1+\frac{1}{\bar Nh} \right)+h^{2} +\tau_m\right). $$
		\item[2.] If  Assumption \ref{asm:cov}(b) holds, then
		$$\|\hat{D}-D\|=O_{P}\left(\frac{1}{\sqrt{n}}\left(1+\frac{1}{\bar Nh} \right)+h^{2}+L_{\varrho_{n}}M_{n}\tau_{m} \right);$$ $$ \|\hat{D}-D\|_{\infty}=O_{P}\left(\frac{\log n}{\sqrt{n}}\left(1+\frac{1}{\bar Nh} \right)+h^{2}+L_{\varrho_{n}}M_{n}\tau_{m} \right), $$
		where  $M_{n}$ is the diverging bound on the processes in  \eqref{eq:Mn} and $L_{\varrho_{n}}$ is the Lipschitz constant in Assumption \ref{asm:gmap} with $\varrho_{n}=1-g(M_{n})$.
	\end{itemize}
\end{corollary}
}
This demonstrates that  the convergence rate of the covariance function results from a combination of a 2-dimensional kernel smoothing rate and the estimation error due to the fact that the transport processes are estimated from data that the underlying distributions generate.  As discussed after Assumption \ref{asm:cov}, 
if $Z$ is sub-Gaussian, then $M_n$ is of the order $\log n$, $M_{n} \sim \log n$. If for example the link function is  $g=(\sqrt{1+4x^2}-1 )/(2x)$, $\varrho_{n} \sim (\log n)^{-1}$ and $L_{\varrho_{n}}\sim (\log n)^k$ for some integer $k$, and if $\tau_m \sim n^{-0.5 +\epsilon}$ for some $\epsilon>0$,  the rate of convergence for $\hat D$ is the same as that for $\hat C$, and the fact that the 
distributions need to be estimated from the data they generate does not affect the convergence in this case. The rate $\tau_m \sim n^{-0.5 +\epsilon}$ is easily achievable, for example when  the minimum number of observations $m$ in \eqref{m} generated by each distribution is of the order $m=m(n) \sim n^{2+\epsilon}$. 

The following central result establishes the $\mathcal{L}^{2}$-convergence rate of $\hat{T}_{i}(t)$, using  cut-off points  $J_n$ as in \eqref{Uhat},
eigenvalues $\lambda_k$ of $C$ as in \eqref{KLC}, $N_i$  and 
$\bar{N}$ as in \eqref{Nbar}, 
and $\kappa$ as in \eqref{eq:T-rescale}.   
\begin{theorem}\label{thm-dis}
	Under Assumptions  \ref{asm:Ti0} - \ref{asm:kernel}, \ref{asm:cov}(a)  and $t_{ij}$ are random samples from $\mathrm{Unif}(0,1)$,  if  $\kappa\geq\|T_{i0} \|_1$ and   
	$\tau_m$ is as in \eqref{tau}, 
	{\small{$$\int  d_{W} \{\hat{T}_{i}(t),T_{i}(t)\}^2\diff t=O_{P}\left( \sum_{k=1}^{J_n}\delta_{k}^{-2}\left\{\frac{\log n}{{n}}\left(1+\frac{1}{\bar N^2h^2} \right)+h^{4} +\tau_{m}^2 \right\}+\frac{J_n}{N_{i}}+\sum_{k=J_n+1}^{\infty}\lambda_{k} \right), $$}}
	where $\delta_{k}=\min_{j\neq k}|\lambda_{j}-\lambda_{k}|$ are the eigengaps.  	
\end{theorem}

The term $\sum_{k=J_n+1}^{\infty}\lambda_{k}$ captures the approximation bias resulting from the finite approximation of the infinite-dimensional eigenexpansion in \eqref{Uhat}, which  decreases as the truncation point  $J_n$ increases. However, as $J_n$ grows, the eigengap $\delta_{J_n}$ approaches zero, making it difficult to distinguish adjacent eigenpairs, counteracting the improvement in approximation error. 
The terms $n^{-1}\{1+(Nh)^{-2}\}$ and $h^{4}$ correspond to the estimation variance and bias of the kernel smoother, while the term $J_n/N_i$ arises from the discrete approximation. Note that $\tau_m$ represents the estimation error of $d_W(\hat{T}_{ij},T_{ij})$, which is negligible if $m=m(n)$ in \eqref{m} diverges sufficiently fast, where  $\tau_m$ is of  the order ${m}^{-1/4}$ or ${m}^{-1/3}$ depending on assumptions and estimation procedures \citep{mull:21:3}, as discussed after Theorem \ref{thm-equclass}. In such cases, $\tau_{m}$ is negligible when $ m\gtrsim n^{2+\epsilon}$ or $ m\gtrsim n^{3/2+\epsilon}$. 

Existing results on representation models for Euclidean functional data only provide the convergence rate of $\|\hat{X}_{i}^{J}-X_{i}^{J}\|$, where $X_{i}^{J}=\sum_{k=1}^{J}\xi_{ik}\phi_{k}$ is the truncated process with a fixed $J$ \citep{yao2005functional}. 
Due to the infinite dimensionality of functional data, obtaining the convergence rate for $\|\hat{X}_{i}-X_{i}\|$ is much more involved; 
 the result in Theorem \ref{thm-dis} appears to be novel even for the much simpler case where processes are Euclidean-valued.  

Phase transitions for estimating  mean and covariance in traditional functional data have been well studied  \citep{cai2011optimal, zhang2016sparse} as measurement designs move from sparse to dense settings. We show in the following that similar results can be obtained for sparsely sampled transport processes. Considering  cases where $\{\lambda_{k}\}_{k=1}^{\infty}$ exhibit polynomial or exponential decay, which are two commonly studied settings for functional data, our main results imply the following corollaries. Here we assume $N_{i}=N$ for all $i=1,\ldots,n$  without loss of generality to simplify notation. In  this case, $\bar{N}=N$.
\begin{corollary}\label{cor-dense}
	Under assumptions in Theorem \ref{thm-dis}, for large enough $ m$ and  $h\asymp (n N^2)^{-1/6}$,
	\begin{itemize}
		\item If $\lambda_{k}\asymp k^{-a}$ with $a>1$, 
		$$\int d_W\{\hat{T}_{i}(t),T_{i}(t)\}^2\diff t=O_{P}\left( J_n^{2a+2}\left\{\frac{\log n}{{n}}+\left(\frac{\log n}{n N^2}  \right)^{2/3} \right\}+\frac{J_n}{N}+J_n^{1-a} \right). $$
		Specifically,  when $(n/\log n)^{a/(3a+1)}/{N} \rightarrow0 $ and $J_n=(n\log n)^{1/(3a+1)}$,
		$$\int d_W\{\hat{T}_{i}(t),T_{i}(t)\}^2\diff t=O_{P}\left( \left(\frac{\log n}{n} \right)^{\frac{a-1}{3a+1}} \right). $$
		\item If $\lambda_{k}\asymp e^{-ck}$ with $c>0$, 
		$$\int d_W\{\hat{T}_{i}(t),T_{i}(t)\}^2\diff t=O_{P}\left( e^{J_n} \left\{\frac{\log n}{{n}}+\left(\frac{\log n}{n N^2}  \right)^{2/3} \right\}+\frac{J_n}{N}+e^{-J_n} \right). $$
		Specifically,  when $(n/\log n)^{1/3}/N\rightarrow0 $ and $J_n\asymp \log(n/\log n )$,
		$$\int d_W\{\hat{T}_{i}(t),T_{i}(t)\}^2\diff t=O_{P}\left( \frac{\log n}{n} \right)^{1/3}. $$
	\end{itemize}
\end{corollary}

According to Corollary \ref{cor-dense}, when the number of observations $N$ is sufficiently large, which refers to the ``ultradense'' case, $J_n/N$ is dominated by  other terms, and the optimal truncation $J_n$ is selected to balance the variance and bias terms. In such cases, the convergence rate of $\int d_W\{\hat{T}_{i}(t),T_{i}(t)\}^2\diff t$ cannot be improved as $N$ increases. However, for  the case where the $N_i=N$  are relatively small but still tend to infinity as $n\rightarrow\infty$, the following holds. 
\begin{corollary}\label{cor-sparse}
	Under assumptions in Theorem \ref{thm-dis}, for large enough $ m$ and $h\asymp (n N^2)^{-1/6}$,
	\begin{itemize}
		\item If $\lambda_{k}\asymp k^{-a}$ with $a>1$,   
		when $N\rightarrow\infty$, $N\lesssim (n\log n)^{a/(3a+1)}$ and $J_n=N^{{1}/{a}}$,
		$$\int d_W\{\hat{T}_{i}(t),T_{i}(t)\}^2\diff t=O_{P}\left( N^{\frac{1-a}{a}} \right). $$
		\item If $\lambda_{k}\asymp e^{-ck}$ with $c>0$, 
		when $N\rightarrow\infty$, $N\lesssim (n/\log n)^{1/3} $ and for a solution  $J_n^{\ast}$  of the equation $\log J_n=\log N-cJ_n,$
		$$\int d_W\{\hat{T}_{i}(t),T_{i}(t)\}^2\diff t=O_{P}\left(\frac{J_n^{\ast}}{N}  \right) =  O_{P}\left(\frac{\log N }{N}  \right) . $$
	\end{itemize}
\end{corollary}
Thus when the $N$ are relatively small but still tend to infinity, the  convergence rate of $\int d_W\{\hat{T}_{i}(t),T_{i}(t)\}^2\diff t$ is dominated by the discrete approximation that results from selecting the optimal infinite truncation point  $J_n$ in \eqref{Uhat}. This is reminiscent of the situation in classical functional data analysis for real-valued random functions, where one may pool data across the sample
when estimating  mean and covariance functions, while such pooling does not apply when predicting individual trajectories.  The  convergence rate for prediction is then determined by the sample size $N$ due to dominance of the approximation error.

Next we consider the sparse case where the numbers of measurements made for each process  $N_{i}$ are strictly finite throughout,  in contrast to the previous result where they are small but diverge, however slowly.  The scenario with fixed $N_i$ reflects designs used in  longitudinal studies,  where distributional data  are sampled at a few random time points for each subject. An example are longitudinal studies in brain imaging where one collects fMRI signals that give rise to connectivity distributions \cp{mull:19:8}; the times when fMRIs are collected are typically very sparse and irregular. 
 As mentioned in Section \ref{sec:meth-spa},  in this sparse case a Gaussianity  assumption needs to be imposed for the latent processes  $Z$ in order to obtain the best linear predictor for estimating $T_i(t)$. 
 For this sparse/longitudinal sampling design, we have the following result for the estimator \eqref{eq:def-T-est-S}.

\begin{theorem}\label{thm-sparse}
	Under Assumptions  \ref{asm:Ti0} to \ref{asm:kernel} and \ref{asm:cov}(b), for the case of finite $N_i\ge 2$ with $\bar{N}= (1/n)\sum_i N_i$, 
	\begin{itemize}
		\item[(a)] $|\hat{\chi}_{il}-\tilde{\chi}_{il}|=O_{P}\left(\frac{\log n}{\sqrt{n}}\left(1+\frac{1}{\bar Nh} \right)+h^{2}+\tau_{m}+L_{\varrho_{n}}M_{n}\tau_{m} \right)  $;
		\item[(b)] For all $i=1,\ldots,n$, $$\int  d_W \{\hat{T}_{i}^{J}(t),\tilde{T}^{J}_{i}(t)\}^2\diff t=O_{P}\left( \frac{\log n}{{n}}\left(1+\frac{1}{\bar N^2h^2} \right)+h^{4}+\tau_{m}^2+L_{\varrho_{n}}^2M_{n}^2\tau_{m}^2 \right), $$
		where $\tilde{T}^{J}_{i}(t)=g(\tilde{Z}_{i}^{J}(t))\odot T_{i0} $ with $\tilde{Z}_{i}^{J}(t)=\sum_{l=1}^{J}\tilde{\chi}_{il}\psi_{l}(t)$, and $M_{n}$ is as in \eqref{eq:Mn}, $\varrho_{n}=1-g(M_{n})$   and $\tau_m$  as in  \eqref{tau}. 
	\end{itemize}
\end{theorem}
We note  that the $\tilde{\chi}_{ij}$ defined in \eqref{bp} are the best linear predictors of the principal component scores of $Z_{i}$ given the data $(Z_{i1},\ldots,Z_{ij})$ and $\tilde{T}^{J}_{i}(t)$ are the transport processes based on these scores $\tilde{\chi}_{ij}$.

\section{Simulations}\label{sec:simu}

We conducted simulation studies to evaluate the numerical performance of the proposed  transport process model \eqref{tm1}, \eqref{tm2}. Trajectories and observed data are generated  as follows.
\begin{itemize}
	\item The underlying process is  $Z_{i}(x)=\sum_{k=1}^{50}\xi_{ik}\phi_{k}(x)$, where $\xi_{ik}\sim N(0,k^{-2})$ and $\phi_{k}(x)=\cos(2(k-1)\pi x)$ for $k>1$ and $\phi_{1}=1$.
	\item The baseline transports $T_{i0} $  correspond to the quantile function of $\mathrm{Beta}(a_{i},b_{i})$, where $a_{i}\sim\mathrm{Unif}(3,4) $ and $b_{i}\sim\mathrm{Unif}(1,2) $. All  $T_{i0} $ are rescaled such that $\|T_{i0}\|_1$ are the same for all $i$. 
	\item The  transport processes are  $T_{i}(t)= U  _{i}(t)\odot T_{i0} $, where $ U  _{i}(t)=g\{Z_{i}(t)\}$ with $g(x)=2\arctan(x)/\pi$.
	\item The measurements are taken at $N$ discrete time points $\{t_{ij}\}_{j=1}^{N}$. The actual observations are random samples $\{{x}_{ijk}:k=1,\ldots,m\}$ from the corresponding distribution of $T_{ij}=T_{i}(t_{ij})$. 
		\item Thus, the observed data are $\{t_{ij}:i=1,\ldots,n;\ j=1,\ldots,N\}$ and $\{{x}_{ijk}:i=1,\ldots,n;\  j=1,\ldots,N;\ k=1,\ldots,m\}$. \end{itemize}

We then applied the proposed method in Section \ref{sec:meth-den}  to predict each $T_i(t)$ based on the transport model  \eqref{tm1}, \eqref{tm2}.
 For each simulation setting, we repeated the procedure 200 times and computed the integrated mean squared error (IMSE) of the reconstruction error as follows:
$$\mathrm{IMSE} = \frac{1}{n}\sum_{i=1}^{n}\int d_{W}\{\hat{T}_{i}(t),{T}_{i}(t)\}^2\diff t,$$
where the integral over $t$ is approximated by a Riemann sum on a dense grid. We considered  both a  random design where $\{t_{ij}\}_{i,j}$ are randomly  sampled from $\mathrm{Unif}(0,1)$ and a fixed  design where the $\{t_{ij}\}_{i,j}$ are equispaced on $(0,1)$.  The results for random design   are in Table  \ref{tab:random}, showing a declining  trend in the IMSE as the sample size $n$, the observations per subject $N$  and the number of observations  $m$ generated by each underlying distribution increase. Moreover, we note that  the IMSE tends to decline more slowly as $n$ increases for a fixed $N$ compared to the situation where  $N$ increases  for a fixed sample size $n$; this is in line with theory. The results for fixed design can be found in  the Supplement. 

\single 

\begin{table}[htbp]
  \centering
  \caption{Monte Carlo averages with standard errors in parentheses of  IMSE based on 200 replications in the sparse design setting.}
    \begin{tabular}{lllllll}
    \toprule
          & \multicolumn{3}{c}{$m=10$} & \multicolumn{3}{c}{$m=200$} \\
    \midrule
          & $n=20$  & $n=50$  & $n=200$ & $n=20$  & $n=50$  & $n=200$ \\
    \midrule
    $N=3$   & 4.04(0.87) & 3.98(0.54) & 3.78(0.53) & 3.30(0.73) & 3.17(0.57) & 2.95(0.49) \\
    \midrule
    $N=5$   & 3.48(0.65) & 3.23(0.49) & 3.09(0.40) & 2.53(0.59) & 2.44(0.49) & 2.30(0.36) \\
    \midrule
    $N=10$  & 2.57(0.49) & 2.52(0.42) & 2.39(0.30) & 1.84(0.42) & 1.79(0.33) & 1.70(0.19) \\
    \midrule
    $N=20$  & 2.05(0.37) & 1.96(0.28) & 1.92(0.18) & 1.37(0.30) & 1.34(0.17) & 1.34(0.12) \\
    \bottomrule
    \end{tabular}%
  \label{tab:random}%
\end{table}%

\double 

\begin{figure}[h]
    \centering
    \begin{subfigure}[b]{0.3\textwidth}
        \centering
        \includegraphics[width=\textwidth]{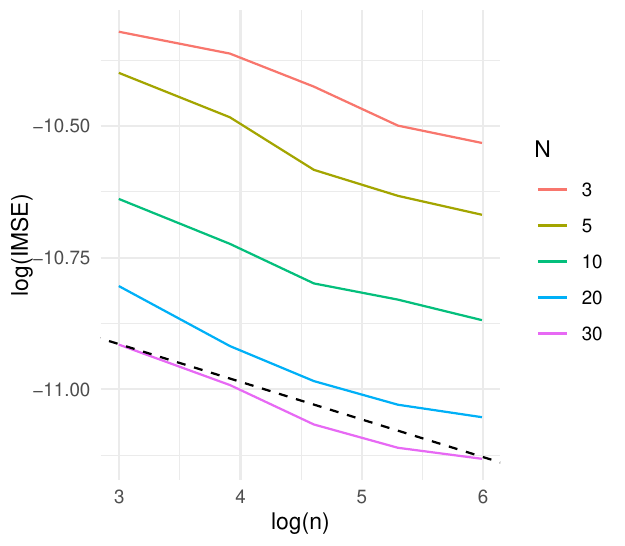}
    \end{subfigure}
    \hfill
    \begin{subfigure}[b]{0.3\textwidth}
        \centering
        \includegraphics[width=\textwidth]{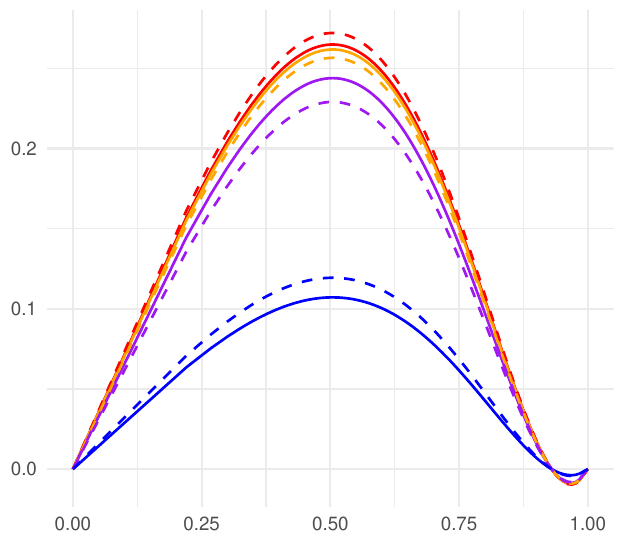}
    \end{subfigure}
    \hfill
    \begin{subfigure}[b]{0.3\textwidth}
        \centering
        \includegraphics[width=\textwidth,height=0.20\textheight]{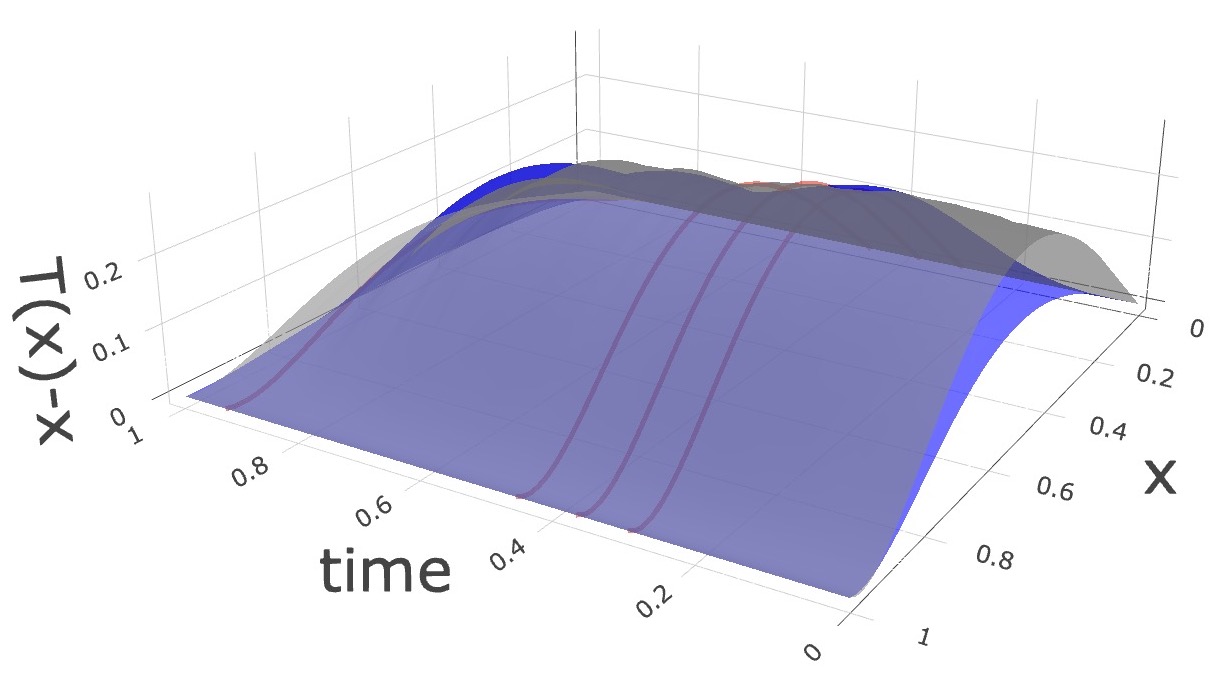}
    \end{subfigure}
    \caption{{\it Left panel:}   $\log(\mathrm{IMSE})$  versus $\log(n)$.  The colored solid lines correspond to different values of $N$. The dashed black line represents the  slope predicted by theory.  {\it Middle panel:}  Sparsely observed transports and corresponding estimates for a randomly selected subject. Different colors represent different time points. The solid lines represent the actual true curves, while the dashed lines stand for their corresponding estimates.  {\it Right panel:} True (gray) and estimated (blue) transport surfaces over time. 
    Red lines indicate  the transports at the observed times,  where  the identity map is subtracted. }
    \label{fig:simu_sparse}
\end{figure}

When the phase transition occurs, Corollary \ref{cor-dense} indicates a proportional relationship such that \(\log(\text{IMSE}) \propto \log(n)\). The left panel of Figure \ref{fig:simu_sparse} illustrates this phenomenon, showing that as \(N\) increases, the relationship between \(\log(\text{IMSE})\) and \(\log(n)\) tends to be linear with the  observed and theory predicted slopes being very close.  The middle and right panels of Figure \ref{fig:simu_sparse} present the observed and estimated transports for the sparse case \((N=4)\) for a randomly selected subject, demonstrating the effectiveness of the  proposed method.

\single 

\suppcontent{
\begin{table}[htbp]
  \centering
  \caption{Monte Carlo averages with standard errors in parentheses for IMSE  based on 200 replications in the dense design setting.}
    \begin{tabular}{clllll}
    \toprule
          &       & $n=50$  & $n=100$ & $n=200$ & $n=400$ \\
    \midrule
    \multirow{3}[2]{*}{$m=10$} & $N=30$  & 1.75(0.33) & 1.71(0.28) & 1.62(0.20) & 1.61(0.17) \\
          & $N=50$  & 1.40(0.16) & 1.36(0.11) & 1.36(0.09) & 1.36(0.07) \\
          & $N=100$ & 1.15(0.09) & 1.14(0.06) & 1.14(0.04) & 1.14(0.03) \\
    \midrule
    \multirow{3}[2]{*}{$m=50$} & $N=3$   & 0.82(0.10) & 0.81(0.07) & 0.81(0.05) & 0.81(0.04) \\
          & $N=5$   & 0.69(0.06) & 0.69(0.04) & 0.69(0.69) & 0.69(0.02) \\
          & $N=10$  & 0.63(0.06) & 0.63(0.04) & 0.63(0.03) & 0.64(0.02) \\
    \midrule
    \multirow{3}[2]{*}{$m=200$} & $N=3$   & 0.59(0.06) & 0.59(0.05) & 0.58(0.03) & 0.59(0.02) \\
          & $N=5$   & 0.52(0.06) & 0.51(0.04) & 0.51(0.03) & 0.52(0.02) \\
          & $N=10$  & 0.50(0.06) & 0.51(0.04) & 0.52(0.03) & 0.53(0.02) \\
    \bottomrule
    \end{tabular}%
  \label{tab:fix}%
\end{table}%
}  

\double

\section{Real data application}\label{sec:appli}
\subsection{Human Mortality}
Human longevity has been actively studied over several decades and analyzing mortality data across countries and calendar years has provided key insights. The Human Mortality Database at www.mortality.org contains yearly age-at-death tables for 38 countries, grouped by age from 0 to 110+. Smooth densities of age-at-death distributions indexed by country and calendar year can be obtained by applying simple smoothing to the  lifetables that are available in this database. We focused on the 33 countries for which data are available for the calendar years from 1983 to 2018. The distributions of age-at-death are  viewed as an i.i.d.  sample of distributional processes   $X_{i}(t)$, where the index $i$ indicates the country and $t$ is calendar year.  

\begin{figure}[H]
  \centering
  \begin{minipage}[b]{0.54\textwidth}
  \centering
    \includegraphics[width=\textwidth]{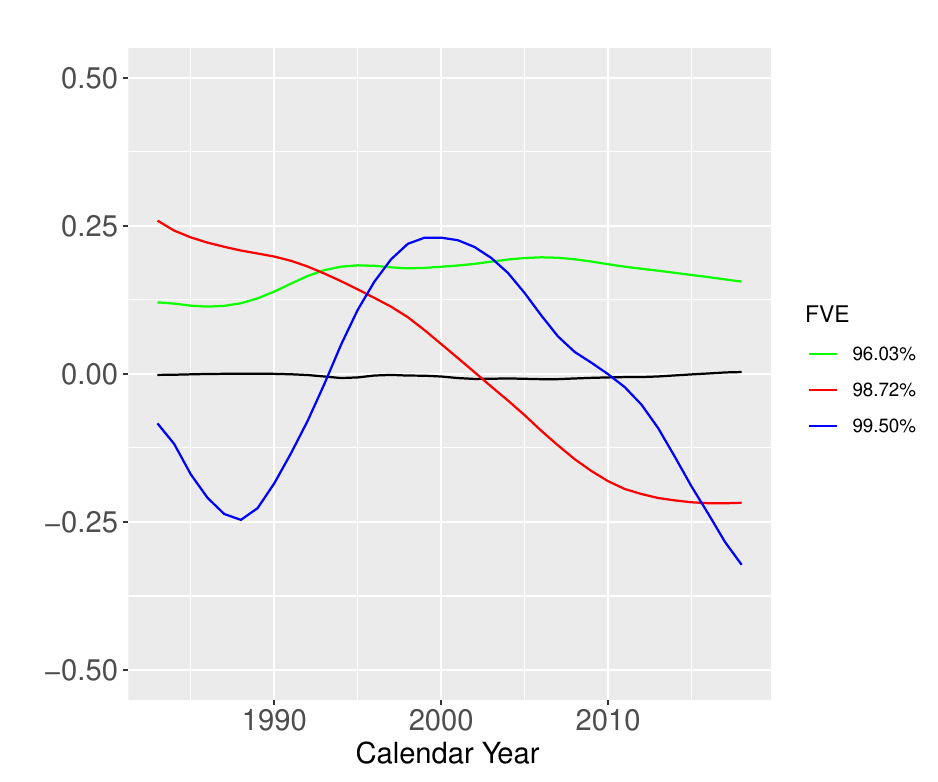}
  \end{minipage}
  \hfill
  \begin{minipage}[b]{0.44\textwidth}
  \centering
    \includegraphics[width=\textwidth]{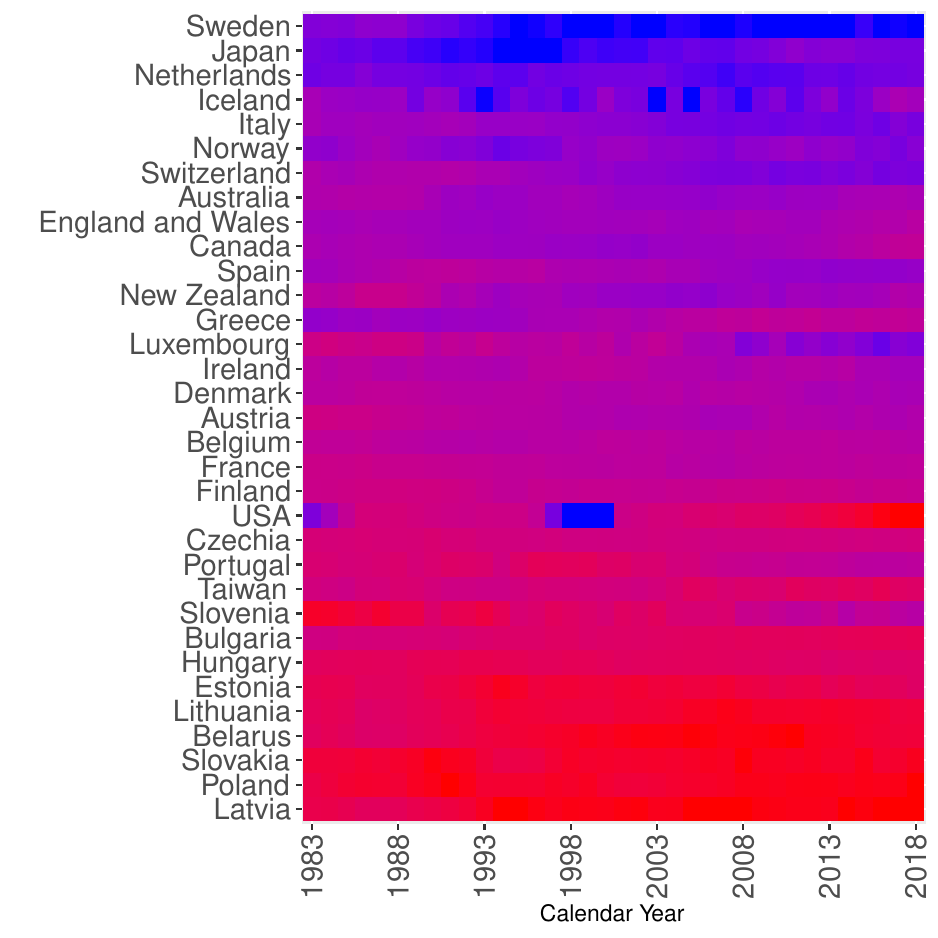}
  \end{minipage}
   \caption{{\it Left panel:}  Estimated mean and eigenfunctions of $ \hat{Z}_{i}(t) $. Black lines represent the mean functions, while the first, second, and third eigenfunctions are depicted in green, red, and blue, respectively. {\it Right panel:}   Transport signs and size of transported mass from the  barycenter to the age distribution for each country and calendar year for males.  Red indicates that transports are predominantly  moving mass to the left (negative sign), and blue that transports are predominantly  moving mass to the right (positive sign). The former is associated with decreased and the latter  with increased longevity when compared to the barycenter of the age-at-death distributions for the corresponding  calendar year. \label{fig:mean_ga}}
\end{figure}

The optimal transports from $\mu_{\oplus}(t)$ to $X_{i}(t)$, denoted by $T_{i}(t)$, form the basis for our analysis.   
The left panel of Figure \ref{fig:mean_ga} shows that the estimated mean functions of the underlying processes \(\hat{Z}_{i}(t_{ij})\) for males, as defined in Equation \eqref{eq:def-hatUZ}, are very close to 0, indicating there is no lack of fit. The estimated mean and eigenfunctions for females, which exhibit similar patterns,  can be found in the Supplement.

\begin{figure}[H]
  \centering
  \begin{minipage}[b]{0.49\textwidth}
  \centering
    \includegraphics[width=0.8\textwidth]{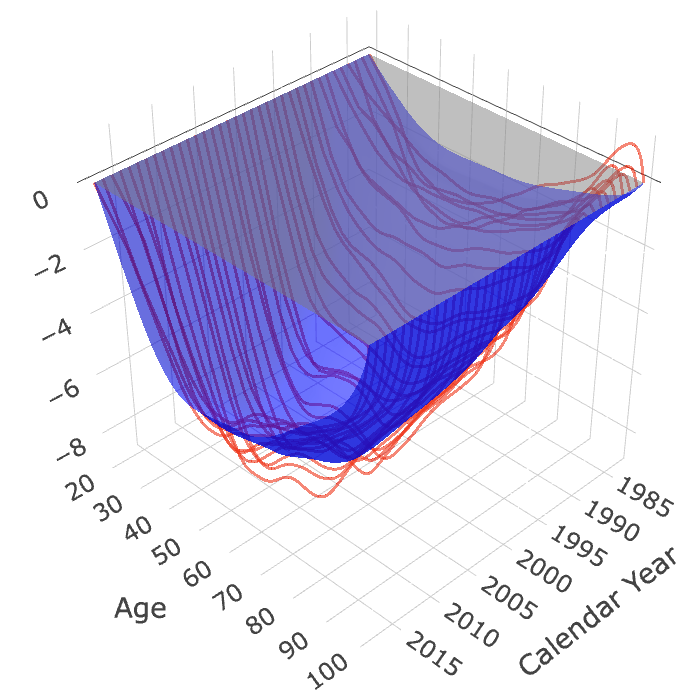}
  \end{minipage}
  \hfill
  \begin{minipage}[b]{0.49\textwidth}
  \centering
    \includegraphics[width=0.8\textwidth]{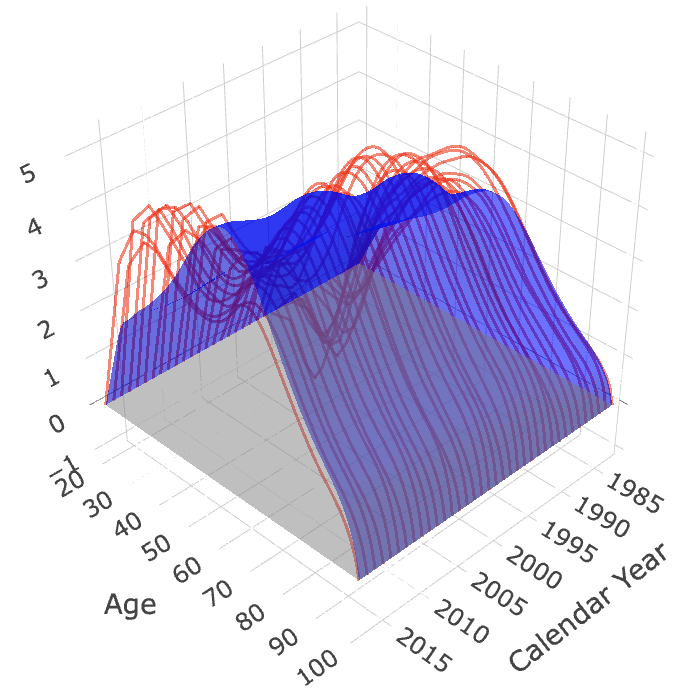}
  \end{minipage}
  \caption{Fitted transport  trajectories obtained form the proposed representations (blue surfaces) for age-at-death distributions for females when using three eigenfunctions for processes $U(t)$, compared with the observed transport processes,  for Belarus (left) where transports have negative signs and indicate  shortened longevity,  and for Spain (right) where they have positive signs and indicate  extended longevity. \label{fig:represent}}
\end{figure}

The representations obtained for two transport process trajectories utilizing   the first three eigenfunctions of two randomly selected countries are shown in Figure \ref{fig:represent}, where we  subtract the identity map for better illustration.  The predicted processes are reasonably close to the data and are seen to provide good fits.

We also explored the sign changes of $T_{i}(t)$ for each country. We found that for most of the richer countries, the signs of $T_{i}(t)$ are positive, meaning that $T_{i}(t)$ moves mass to the right from the Fr\'echet mean, associated with delayed age-at-death and increased longevity, while the  $T_{i}(t)$ with negative signs are primarily associated with lower income countries, where longevity is below average. But there are also interesting sign changes throughout the calendar period for various countries. The right panel of Figure \ref{fig:mean_ga} shows the signs and the amount of mass transported to the right or left for males. Similar results for females can be found in the Supplement.

\single

\double


\subsection{Distributional processes  of glucose concentrations}
Diabetes is a growing global health concern. Continuous glucose monitoring  devices provide detailed tracking of glucose levels at regular intervals over extended periods. We utilize glucose concentration data from \cite{hall2018glucotypes} to obtain daily distributions of glucose levels for a sample of $n=52$ subjects, for whom glucose levels (in mg/dL)  were recorded 
every 5 minutes over a variable time range of 4 to 10 days after the initial recording. For each day where recordings are available,  we applied a kernel density estimator to derive the smoothed density function based on the observed glucose levels, similar to an approach of 
\cite{mata:21}. 
 Due to  individually varying recording periods across  subjects, the resulting  densities  $\mu_{ij}$, where $i=1,\dots,n$ is the subject index  and $j$ a  measurement index, indicating the $j$-th day in the study, are sparsely observed in time. 

\single

\double

Implementing the proposed approach, the variable of interest is the optimal transport \( T_{ij} \) from \( \hat{\mu}_{\oplus}(t_{ij}) \) to \( \mu_{ij} \), where \( \hat{\mu}_{\oplus}(t) \) is the estimated Fr\'echet mean of \( \{\mu_{ij}\}_{i,j} \) at time \( t \). The transport \( T_{ij} \) reflects how the \( i \)-th subject compares with the population at time \( t_{ij} \), which is important for understanding the personal nature of glucose regulation. Due to the sparse nature of the measurements \( t_{ij} \), we utilize Fr\'echet regression \citep{petersen2019frechet} to estimate \( \hat{\mu}_{\oplus}(t) \). To illustrate the proposed approach for the case of distributional processes that are observed sparsely in time, we randomly select two days as test data for subjects with more than four days of measurements to evaluate the prediction accuracy of the proposed method,  using the remaining data as training data. 

\begin{figure}
  \centering
  \includegraphics[width=0.95\textwidth]{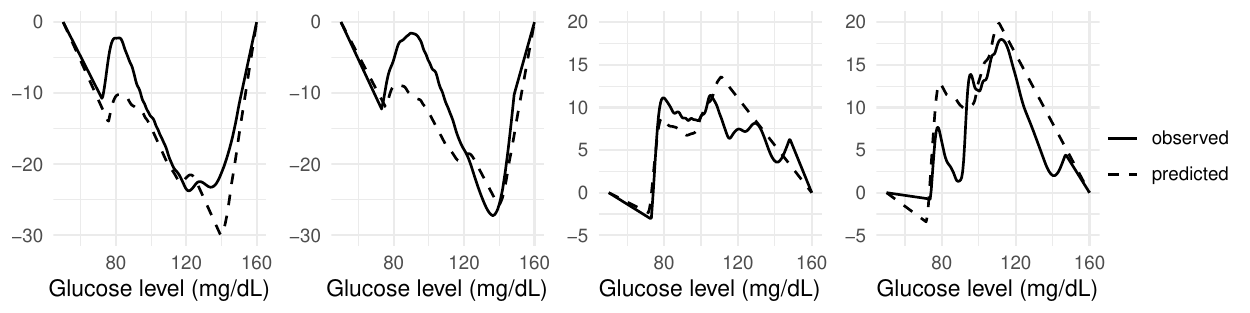}
  \caption{Observed transports in the testing set (black lines) and predicted (dashed lines) transports of glucose density data for two randomly selected subjects. These transports are illustrated by subtracting the identity map for clarity. The left two panels represent the data for a randomly selected  subject at times \( t=1 \) and \( t=2 \), while the right two panels represent  the data for another subject at times \( t=5 \) and \( t=6 \).\label{fig:gluco}} 
\end{figure}

Figure \ref{fig:gluco} depicts  the observed transports for the test data and the estimated transports, demonstrating that the proposed method  predicts the transports reasonably well even in the presence of  sparse observations. For the subject corresponding to the left two panels of Figure \ref{fig:gluco}), the transports are below the identity map, indicating that this subject's glucose concentration levels are below average at these two times. For the  subject corresponding to the right  two panels of Figure \ref{fig:gluco}), the transports are above the identity map and the transport mass is larger at \( t=6 \) compared to \( t=5 \). This indicates that the glucose concentration levels for this subject are slightly above average at \( t=5 \) and further above average  at \( t=6 \), indicating a worsening situation. 


The estimated mean and eigenfunctions of the underlying Gaussian processes are presented in Supplement, where it is found that  the first three eigenfunctions explain $99.85\%$ of the variance.

\single

\double

\section{Concluding Remarks}


Modeling object valued data has found increasing interest in recent years  \citep{marron2021object}. A key challenge in this context is the lack of linear structure, which plays an essential role in classical statistics and especially in principal component analysis. One 
approach to surmount this obstacle involves mapping the data into a linear space, however this remains unsatisfactory since for maps that are isometric the  inverse map is only defined on a subset of the image space; a typical  example is local linearization with tangent bundles \cp{bigo:17}.  
 Direct linearizing transforms are generally invertible on the entire space  \citep{petersen2016functional}  
but are not isometric and lead to metric distortions.  All of this suggests that  intrinsic representations  are an attractive alternative, where the analysis takes place entirely in the given metric space and therefore also may allow for better interpretation.  

Converting  distribution-valued processes  to transport processes, as we propose here, has two main benefits. First, the transformation to transport processes is isometric, and we can equate the analysis of transport processes to that of  distribution processes. Second, this transformation provides a centering operation for distribution-valued processes, effectively overcoming the absence of a subtraction operation, as  transport processes are automatically centered and their mean is the identity.

A central tenet of our proposed model is the decomposition of the time-varying transport process into a real-valued stochastic process and a random transport that characterizes the entire transport trajectory. 
As demonstrated in Section \ref{sec:appli}, the proposed representation model and decomposition works well for real-world data. The decomposition is facilitated by a multiplication operation between a scalar and a transport map \citep{mull:21:3}, which we use to exploit  an equivalence relation within the transport space. Consequently, the transports $T(t)$ reside in an equivalence class, providing the geometric basis for the proposed representation. The stochastic process part in this decomposition introduces a real-valued stochastic process with ensuing eigenrepresentation.  This is a major advantage 
as one then can bring to bear many concepts of 
functional data analysis, most importantly functional principal component analysis, overcoming  the fact  that there is no linear structure in the distribution space. When the latent real-valued stochastic process can be considered to be Gaussian, the existing machinery that connects functional data analysis for the sparsely sampled case with longitudinal data analysis can be brought to bear.  

\bco 

A further advantage of considering transport processes is their capacity to model multivariate distribution processes. Specifically, in scenarios where ${X(t),Y(t)}$ constitute a pair of distributional trajectories and the relationship between these two components is of interest, one can consider  optimal transports $T(t)$ from $X(t)$ to $Y(t)$, which represent geodesics in the Wasserstein space. This connection
adds to the appeal of transport processes. Furthermore, while we provide a detailed development here for the case of distribution-valued processes, this 
can serve as a blueprint for a larger class of metric-space valued processes in unique geodesic spaces where transports can be considered to move random objects along geodesics \cp{mull:23}. Further development along these lines is left for future research. 

\fi

%
%
%
%
%

\bibliographystyle{agsm}

\bibliography{report}
\end{document}